\documentclass[twocolumn]{aastex701}

\usepackage[utf8]{inputenc}
\usepackage{newunicodechar} 
\newunicodechar{✱}{\ensuremath{*}}
\usepackage{pifont}
\usepackage{needspace}
\usepackage{amsmath} 

\usepackage{changepage}
\usepackage{float}

\newcommand{\mjup}{\ensuremath{\mathrm{M_{jup}}}}

\newcommand{\nrc}{NRC Herzberg Astronomy and Astrophysics,
5071 West Saanich Road,
Victoria, BC, V9E 2E7, Canada}
\newcommand{\ucsd}{Department of Astronomy \& Astrophysics,  University of California, San Diego, La Jolla, CA 92093, USA}
\newcommand{\nmsu}{Department of Astronomy, New Mexico State University, 1320 Frenger Mall, Las Cruces, NM 88003, USA}
\newcommand{\uvic}{Department of Physics and Astronomy, University of Victoria,
3800 Finnerty Road, Elliott Building,
Victoria, BC, V8P 5C2, Canada}
\newcommand{\mac}{Department of Physics \& Astronomy, McMaster University, 1280 Main Street West, Hamilton, ON, L8S 4L8, Canada}

\newcommand{\ucsc}{Department of Astronomy \& Astrophysics, University of California, Santa Cruz, CA 95064, USA}
\newcommand{\ucsb}{Department of Physics, University of California, Santa Barbara, CA 93106, USA}
\newcommand{\ucla}{Department of Earth, Planetary, and Space Sciences, University of California, Los Angeles, CA 90095, USA}

\begin{document}

\title{Detecting and Characterizing Companions with a Calibrated Gaia DR2, DR3, and Hipparcos Catalog (G23H)}

\author[0000-0001-5684-4593]{William Thompson}
\affiliation{\nrc}
\email[show]{william.thompson@nrc-cnrc.gc.ca}

\author[0000-0001-9582-4261]{Dori Blakely}
\affiliation{\uvic}
\affiliation{\nrc}
\email{dblakely@uvic.ca}  

\author[0000-0002-6618-1137]{Jerry W. Xuan}
\altaffiliation{Heising-Simons Foundation 51 Pegasi b Fellow}
\affiliation{\ucla}
\email{jerryxuan@g.ucla.edu}  

\author[0000-0002-9820-1884]{Simon Blouin}
\affiliation{\uvic}
\email{sblouin@uvic.ca}

\author[0000-0002-2696-2406]{Jingwen Zhang}
\affiliation{\ucsb}
\email{jwzhang@ucsb.edu}

\author[0000-0002-6773-459X]{Doug Johnstone}
\affiliation{\nrc}
\affiliation{\uvic} 
\email{Douglas.Johnstone@nrc-cnrc.gc.ca}

\author[0000-0003-2233-4821]{Jean-Baptiste Ruffio}
\affiliation{\ucsd} 
\email{jruffio@ucsd.edu}

\author[0000-0001-6975-9056]{Eric Nielsen}
\affiliation{\nmsu}
\email{nielsen@nmsu.edu}

\author[0000-0003-3430-3889]{Jessica Speedie}
\altaffiliation{Heising-Simons Foundation 51 Pegasi b Fellow}
\affiliation{\uvic}
\affiliation{Department of Earth, Atmospheric, and Planetary Sciences, Massachusetts Institute of Technology, Cambridge, MA 02139, USA}
\email{jspeedie@mit.edu}

\author[0000-0003-2649-2288]{Brendan P. Bowler}
\affiliation{\ucsb}
\email{bpbowler@ucsb.edu}

\author[0000-0001-7402-8506]{Alexandre Bouchard-Côté}
\affiliation{Department of Statistics, University of British Columbia, Vancouver, BC V6T 1Z4, Canada}
\email{bouchard@stat.ubc.ca}

\author[0000-0003-4557-414X]{Kyle Franson}
\altaffiliation{NHFP Sagan Fellow}
\affiliation{\ucsc}
\email{kfranson@ucsc.edu}

\author[0000-0002-3199-2888]{Sarah Blunt}
\affiliation{\ucsc}  
\email{sarah.blunt.3@gmail.com}

\author[0009-0008-9687-1877]{William Roberson}
\affiliation{\nmsu}  
\email{wcr@nmsu.edu}

\author[0000-0001-5383-9393]{Ryan Cloutier}
\affiliation{\mac}
\email{ryan.cloutier@mcmaster.ca}

\author[0009-0007-6766-2040]{Andre Fogal}
\affiliation{\uvic}
\email{afogal@uvic.ca}

\author[0009-0009-2223-2404]{Kaitlyn Hessel}
\affiliation{\uvic}
\email{khessel@uvic.ca}

\author[0000-0002-4164-4182]{Christian Marois}
\affiliation{\nrc}
\affiliation{\uvic}
\email{Christian.Marois@nrc-cnrc.gc.ca}

\author[0009-0008-1229-3230]{Alexandra Rochon}
\affiliation{\mac}
\email{rochoa3@mcmaster.ca}

\newcommand{\jwz}[1]{\textcolor{magenta}{#1}}

\begin{abstract}

The anticipated release of Gaia DR4 epoch astrometry promises to enable the detection of thousands of exoplanets through the astrometric motion method. Here, we present a composite catalog and modeling framework that extracts the maximum information from existing Hipparcos and Gaia data releases in advance of DR4. We calibrate Gaia DR2 proper motions and DR3–DR2 scaled position differences against the Gaia DR3 reference frame, and combine these with the Hipparcos-Gaia Catalog of Accelerations, the Hipparcos intermediate astrometric data, Gaia astrometric excess noise, and Gaia radial velocity variability constraints. We implement a joint likelihood model for these data in the orbit-fitting code Octofitter that marginalizes over Gaia's unpublished observation epochs. This results in full orbit posteriors that can be computed uniformly for a large class of companions. We compare these posteriors to published orbital solutions for 25 stellar binaries from the Sixth Catalog of Orbits of Visual Binary Stars, recovering all companions at high significance and broadly consistent orbital separations. 
We then recover independent evidence to support the existence of 94 of 120 tested Jovian exoplanetary systems known from the NASA Exoplanet Archive (in addition to 3 known stellar companions, and one previously detected planet we now rule out). We demonstrate that in some cases like 14 Her b, the posteriors are sufficiently concentrated to confirm the planetary nature of a signal using only existing published data from Gaia and Hipparcos. We also find no false positives among a sample of 25 RV-quiet standard stars without significant Hipparcos-Gaia accelerations. Our method can in some cases break degeneracies inherent to proper motion anomaly or excess noise modeling alone by resolving orbital curvature within the Gaia baseline, enabling us to rule out both inner spectroscopic binaries and wide stellar companions in many cases. The catalog of calibrated DR2 and DR3 data for 18 million stars and updated version of Octofitter are made publicly available for the community.

\end{abstract}


\section{Introduction}

Of the numerous methods developed to detect exoplanets, the astrometric motion method pioneered more than 60 years ago \citep[e.g.][]{vandeKamp1963} is now poised, thanks to Gaia, to become a primary way to detect large planets on moderate separation orbits. The anticipated release of Gaia DR4 (hereafter GDR4) will provide epoch astrometry ---the individual astrometric measurements used to fit stellar positions, parallaxes, and proper motions---for billions of stars, over a time span of 5.5 years. These measurements are expected to be sensitive enough to detect thousands of planets \citep{Lammers2026}.

Even without Gaia epoch astrometry, derived parameters from the Gaia DR2 \citep[][hereafter GDR2]{gaia_dr2} and now DR3 \citep[][hereafter GDR3]{gaia_dr3} catalogs have proven useful, when combined with Hipparcos \citep{hipparcos_1} measurements from 20 years prior, as indicators that particular stars have astrometric accelerations consistent with stellar or sub-stellar companions. Derived catalogs that cross-match and calibrate the Hipparcos reference frame against Gaia's include \citet{kervella_stellar_2019} and \citet[the Hipparcos-Gaia Catalog of Accelerations, or HGCA]{Brandt2019}. The ``proper motion anomaly,'' or deviation from long-term straight line motion, reported in these works have led directly to a number of planetary detections though either imaging or radial velocity follow-up, and have been widely used in the determination of dynamical masses of widely orbiting massive companions after they are detected \citep[e.g.][]{Snellen2018, Brandt2019, Brandt2021, Xuan2020, Xuan2020b, Venner2021, currie_2023,  Xiao2024, Zhang2024, Zhang2025, An2025}.


By reporting the individual epoch measurements of each star, DR4 is expected to move the astrometry technique beyond indicators of planetary companions and mass constraints to the independent detection and characterization of individual systems and the study population demographics on a grand scale. In anticipation of this scientific potential, this work examines what additional constraints can be achieved today using the existing, previously published catalogs from Hipparcos, Gaia DR2 (hereafter GDR2), and Gaia DR3 (hereafter GDR3).

\begin{figure*}
\centering
\includegraphics{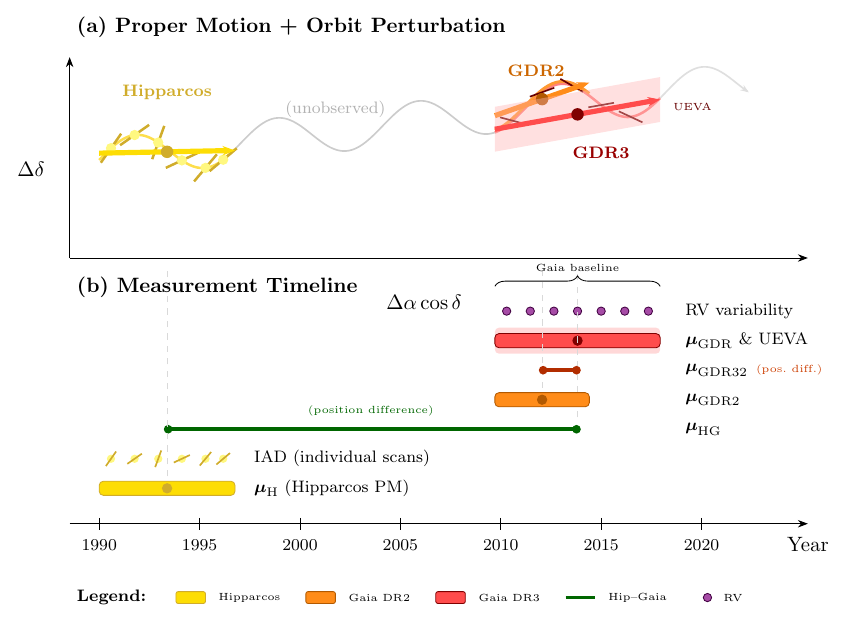}
\caption{Conceptual schematic (not to scale) showing how the different datasets correspond to different time ranges. Linear sky motion is shown without the addition of annual parallax. The proper motions for each data source result from published linear fits to the perturbed motion over different time windows \citep{kervella_stellar_2019,hgca_dr2}. The UEVA Gaia astrometric excess noise constraint is deduced from published uncertainty inflation applied within the GDR3 linear fit \citep{pmex1}. The RV variability is deduced from the uncertainty in the published mean Gaia RV of the star, itself calculated from the sample variance over the GDR3 baseline \citep{paired} The two scaled position differences $\mathbf{\mu_{\rm HG}}$ and $\mathbf{\mu_{\rm DR32}}$ are calculated from the displacement over the time difference between the Hipparcos and GDR3 positions (sourced from the HGCA), and GDR2 and GDR3 positions respectively. The IAD is the only source where each individual measurement is available for modeling. The net proper motion and positions of the IAD are marginalized out to avoid double-counting with the HGCA.\label{fig:schematic}}
\end{figure*}

In this paper, we develop a method that combines several approaches described in the literature:
\begin{itemize}
    \item Gaia-Hipparcos proper motion anomaly \citep{Brandt2019,kervella_stellar_2019};
    \item Hipparcos intermediate astrometry data (IAD) modeling \citep{hipparcos_2, nielsen_iad};
    \item Gaia DR3 vs DR2 proper motion anomaly \citep[incl.][]{Feng2024, thompson_eps_eri, Ribas2025};
    \item Gaia radial velocity (RV) variability \citep{paired}; 
    \item Gaia astrometric excess noise modeling \citep{pmex1}.
\end{itemize}
A schematic visualizing the different measurements is shown in Figure \ref{fig:schematic}.

We offer improvements to several of these methods, including estimates and uncertainties for GDR3 astrometric noise and a calibration that places GDR2 proper motions and positions in the GDR3 reference frame. We then show how these methods can be integrated into an efficient MCMC framework that robustly marginalizes over uncertainties in calibration parameters and the (as-yet) unpublished Gaia sampling epochs.

Our results provide the tightest constraints yet on the presence or absence of planetary companions to millions of nearby Gaia stars. For stars in the Hipparcos catalog, the constraints are sometimes sufficient to independently detect or confirm Jovian planets on $\sim 3-20$ AU orbits. We demonstrate the recovery of 94 out of the total 120 systems in the NASA Exoplanet Archive \citep[hereafter, NEA;][]{Christiansen2025} within 40 pc, with planets with periods greater than 1 year, and orbiting single-stars in the Hipparcos catalog. Several of these are can be placed conclusively in the planetary regime through Gaia and Hipparcos data alone.

We provide a public data catalog\footnote{\href{http://dx.doi.org/10.11570/26.0002}{http://dx.doi.org/10.11570/26.0002}}
which includes our calibrated GDR2 proper motions, GDR3-GDR2 scaled position differences, Gaia noise estimates for each source, and collates information from GDR3, the HGCA, and the \texttt{paired} catalog. Combined with an update to the comprehensive orbit modeling package Octofitter \citep{octofitter}, this allows the method to be run on almost any Gaia star within 1 kpc brighter than 16th magnitude (the volume and magnitude cuts applied when generating this catalog). This package also allows these constraints to be freely combined with other exoplanet data: images, radial velocities, relative astrometry, and so on.

\vspace{1em}
The paper is organized as follows. In Section \ref{sec:catalog-construction} we begin by presenting our calibration of GDR2 proper motions and the GDR3-GDR2 scaled position difference in the GDR3 reference frame. Next, we derive a revised version of the Gaia noise calibration maps originally developed by \citet{pmex1}. 
We then add estimates of Gaia's per-transit RV uncertainty from the \citet{paired} catalog.

In Section \ref{sec:modelling}, we discuss our approach to modeling the Hipparcos, GDR2, and GDR3 proper motions; the Hipparcos IAD; the GDR3 astrometric excess noise; and the GDR3 RV excess noise in the open-source orbit modeling code Octofitter.

With this catalog and this orbit modeling code in hand, in Section \ref{sec:results} we test this method on a sample of confirmed exoplanetary systems, finding good agreement with published solutions. We finally test the method on a set of stars with quiet, long-basline archival RVs to test for false positives.

In Section \ref{sec:discussion}, we compare our approach with that of \citet{pmex1} and list important caveats. Finally, we conclude in Section \ref{sec:conclusion}.

\section{Catalog Construction}
\label{sec:catalog-construction}

We aim to construct a catalog of proper motions drawn from Hipparcos, GDR2, and GDR3 where each data point is placed into the same reference frame, namely that of GDR3.
To do so, we follow a very similar procedure to that used in \citet{hgca_dr2} and in \citet{hgca_dr3}. 


\subsection{Hipparcos-Gaia Proper Motion Anomaly}\label{sec:hgca}

To calibrate Hipparcos proper motions and Hipparcos-Gaia scaled position differences against the Gaia DR3 velocity reference frame, we followed the procedures laid out in \citet{hgca_dr2} and \citet{hgca_dr3}. We replicated the results of the DR3 version of the HGCA within uncertainties for most stars. Satisfied with the agreement, we elected to exactly adopt the HGCA values in this catalog rather than complicate the literature. As a result, users of this catalog and code can exactly replicate a fit to the HGCA velocity subset by simply dropping the other epochs. In all cases when Hipparcos proper motions or Hipparcos-Gaia position differences are used, the GDR3 version of the HGCA \citep{hgca_dr3} should be cited prominently.

\subsection{Calibration of Gaia DR2 vs.\ DR3 velocities}\label{sec:cal-dr2-dr3}

When stars accelerate, the proper motion measured over a shorter baseline does not in general match the proper motion measured over a longer baseline. GDR2 measured and reported the best-fitting straight-line velocity over the first 22 months of data, while GDR3 reports the same calculation over 34 months of data. A difference between GDR2 and GDR3 is expected for both wide-companions imparting a steady acceleration on the primary and closer in companions where GDR2 will fit a straight line to fewer or a less complete orbital cycle than GDR3. If these perturbations, the Gaia measurement epochs, and fitting process are all modeled carefully, these differences provide a proper motion anomaly (PMA) from entirely within Gaia's measurement baseline. 

This approach of comparing DR3 and DR2 reductions has been used in the literature, in particular by \citet{Feng2024} and related works.

A barrier to using the GDR2 vs.\ GDR3 velocities is that the proper motion reference frame changed significantly between GDR2 and GDR3, globally, as a function of position, and as a function of magnitude \citep{lindegren_gaia_frame,cantat_gaudin_brandt}. Some works such as \citet{Ribas2025} have directly compared the raw GDR2 and GDR3 proper motions, ignoring these differences, which could lead one to infer a significant acceleration where none exists.
In works such as \citet{Feng2022}, the authors adopt global calibration measures, and then use flexible offset and jitter parameters on the GDR2 vs. GDR3 reference frame shift, but these latter local adjustments are handled only at the individual star level. 

Here, we take a different approach inspired by the HGCA: global, magnitude-dependent, and local calibrations calibration of the GDR2 measurements against the reference frame of GDR3. This approach pools information from large numbers of stars to produce calibrated GDR2 proper motions that can be directly compared with those of GDR3.

To calibrate the GDR2 proper motions and GDR3-GDR2 scaled position differences, we begin by selecting from GDR2 and from GDR3 all stars brighter than 16th magnitude within 1 kpc. 
Since the source identifiers in GDR3 may not correspond to the same source identifier in GDR2, we cross match the two catalogs using the precomputed GDR3 \verb|dr2_neighbourhood| table, taking the first matching source ID. In general, the correspondence between GDR2 and GDR3 sources is not one-to-one, so this catalog should be used with caution on stars with multiple matches in that table.

We inflate all proper motion uncertainties in the GDR3 catalog by the global 1.37 factor found in \citet{hgca_dr3} to account for a known underestimate of the statistical uncertainties.

Next, we propagate the astrometry of both catalogs to their central measurement epochs following the same approach as \citet{hgca_dr2, hgca_dr3}. This removes (or substantially reduces) the correlation between the catalogs' position measurements and their proper motions such that it can be neglected in future calculations.

For subsequent calculations, we must estimate the correlation between any given GDR2 and GDR3 parameter (e.g.\ proper motion in R.A.). We  follow \citet{thompson_eps_eri} by approximating this as $$\rho_{dr2,dr3}=\sqrt{\min(\frac{N_{AL,DR2}}{N_{AL,DR3}},\frac{N_{AL,DR3}}{N_{AL,DR2})}},\label{eq:rho}$$ where $N_{AL,DR2}$ and $N_{AL,DR3}$ are the number of along scan measurements matched in the GDR2 and GDR3 reductions respectively. This approximately accounts for the measurements shared between the two catalogs\footnote{Note that for a small fraction of sources, GDR2 contains more matched measurements than GDR3}.

Throughout the modeling of this catalog, we account for the correlation between GDR2 and GDR3 values which acts to \emph{increase} the significance of any observed change in position or proper motion due to the shared data. This is because of the third term in the formula for calculating the variance of the difference between two positively correlated variables:
$\sigma^2 = \sigma_{dr2}^2 + \sigma_{dr3}^2 - \rho_{dr2,dr3}\sigma_{dr2}\sigma_{dr3}$. A positive $\rho_{dr2,dr3}$ reduces the variance. This effect may be familiar due to its use in correlated double-sampling, which is used to improve detector noise in other astronomy domains.

\begin{figure*}
    \includegraphics[width=\linewidth]{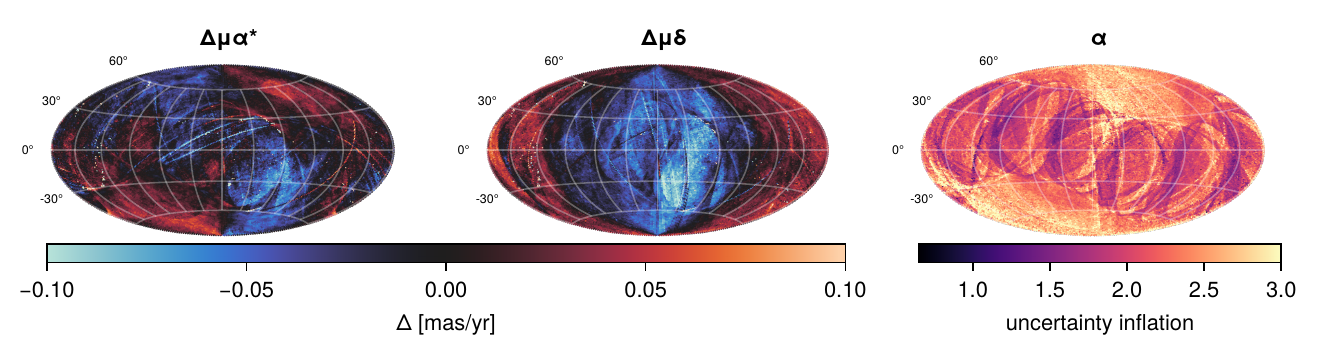}
    \caption{Correction applied to the GDR2 proper motions and uncertainties (Section \ref{sec:cal-dr2-dr3}). The colors indicate the median correction within each Healpix level 6 bin.}
    \label{fig:dr2-cal}
\end{figure*}

\begin{figure}
    \centering
    \includegraphics[width=\linewidth]{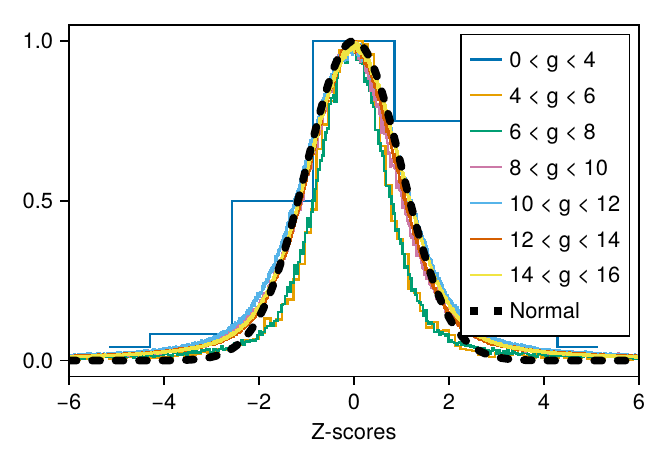}
    \caption{Z-scores of corrected GDR2 proper motions compared to GDR3.}
    \label{fig:zscores-dr2}
\end{figure}

To calibrate GDR2 against GDR3, we begin by addressing the global frame rotation \citep{lindegren_gaia_frame} in GDR2 compared to GDR3.
We perform an initial global, magnitude-dependent rotation fit to the reference frames between GDR2 and GDR3. The magnitude dependence of this fit is motivated by \cite{cantat_gaudin_brandt}. 

We use 84 magnitude bins with the following spacings:
\begin{itemize}
    \item one bin from 0--3 mag;
    \item bins of 0.25 mag from 3 to 10.5;
    \item bins of 0.10 mag from 10.5 to 12.5;
    \item bins of 0.05 mag from 12.5 to 13.5;
    \item bins of 0.25 mag from 13.5 to 16 mag.
\end{itemize}

We only use one bin for magnitudes 0--3 since the Gaia catalogs have very few bright stars due to detector saturation.
Within each magnitude bin, we fit the proper motions from DR2 against DR3 as a global rotation. We perform this fit using three iterations of a weighted least squares, where $\geq 3\sigma$ outliers are rejected each time, and $\sigma$ is calculated as 1.4826 times the median absolute deviation. 

We formulate covariance matrices for the GDR2 and GDR3 R.A.\ and Dec.\ velocities accounting for both the R.A./Dec.\ covariance in each centrally-propagated catalog, and the estimated correlation between the two catalogs.

Next, we proceed with a local calibration of the GDR2 proper motions. We divide the sky into equal area HEALPix level 6 bins \citep{healpix}. Within each bin, we ignore proper motion outliers that are $\geq 10\sigma$ according to either their stated R.A. or Dec. uncertainties, after accounting for both the GDR2 and GDR3 uncertainties and their correlation.
We then fit a Gaussian mixture model to account for proper motion offsets and uncertainty inflation of the same form as used by \citet{hgca_dr2}.
We fit for a shift in the  R.A.\ and Dec.\ proper motions and an error inflation term $\alpha$. 
Each source is considered to be drawn from one of two distributions, one which matches the stated uncertainties (inflated by $\alpha$) and one which represents outliers, having a Gaussian distribution of 1.0 mas/yr. The prior between the two distributions for each point is taken to be 50\%, resulting in the closed form solution presented in \cite{hgca_dr2}.
We noted a very small number of local healpix bins where $\alpha$ converged to a very small value. In these cases, we replaced $\alpha$ for that bin with that of the previous (adjacent) HEALPix bin.

After this local calibration of the GDR2 proper motions, we address a magnitude-dependent underestimate of the GDR2 proper motion uncertainties that is not captured by our local correction (which is dominated by faint stars).
We apply the same mixture model globally, fitting only for an uncertainty inflation term $\alpha$. For this step, we used magnitude bins of $(0,3), [3, 5), [5, 7),$ and $(7,9]$.

Our corrections applied to GDR2 are visualized in Figure \ref{fig:dr2-cal}. There are several strong systematic patterns visible: a known global rotation, fine structure at the level of the scan law, and significant uncertainty inflation that corresponds almost exactly to the GDR2 scan law.

Figure \ref{fig:zscores-dr2} plots the Z-scores (the errors in units of standard deviation) between the calibrated DR2 proper motions and their counterparts from DR3, after accounting for the approximate correlation between the two catalogs. We find as desired that the calibrated proper motions follow a Gaussian distribution with long tails expected from accelerating stars. Stars brighter than 4th magnitude are less reliable (see Section \ref{caveats}). Stars between 4th and 8th magnitude appear somewhat under-dispersed, indicating that their uncertainties are somewhat overestimated.


\begin{figure*}
    \includegraphics[width=\linewidth]{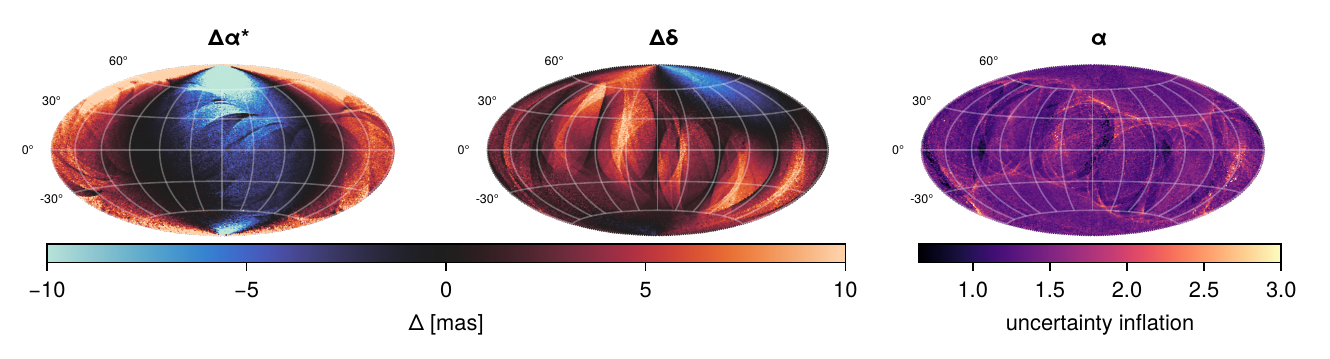}
    \caption{Correction applied to the GDR3--GDR2 scaled position differences (Section \ref{sec:cal-dr32}). The colors indicate the median correction within each Healpix level 6 bin.}
    \label{fig:dr32-cal}
\end{figure*}

\begin{figure}
    \centering
    \includegraphics[width=\linewidth]{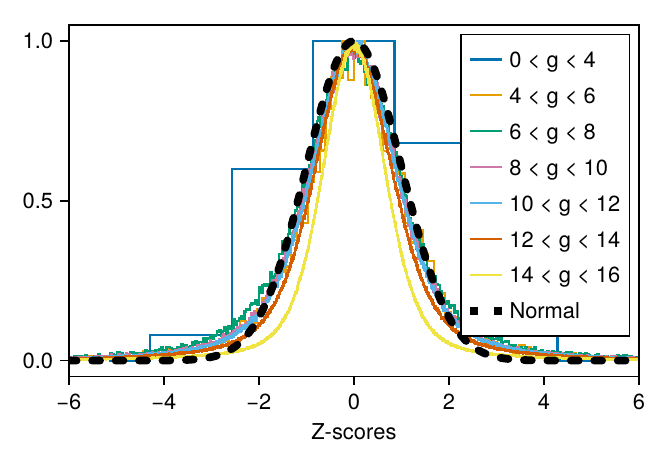}
    \caption{Z-scores of corrected GDR3--GDR2 scaled position differences compared to GDR3 proper motions.}
    \label{fig:zscores-dr32}
\end{figure}

\subsection{Calibration of GDR3--GDR2 scaled position differences}\label{sec:cal-dr32}

Like the G-H proper motion reported by the HGCA, we can calculate an effective proper motion using the scaled position difference between GDR3 and GDR2. Whereas H-G from the HGCA represents the long-term proper motion, GDR3-GDR2 is sensitive to very short term changes in proper motion, since the centrally-propagated measurement epochs for GDR2 and GDR3 often differ by only a handful of months.

To calibrate the positions in GDR2 against their counterparts in GDR3, we follow a similar procedure to that of Section \ref{sec:cal-dr2-dr3}. 

We start by performing an iterative global fit of frame rotation between the two set of catalog positions, after accounting for the GDR3 proper motions between their central epochs. We use the same 84 magnitude bins used in Section \ref{sec:cal-dr2-dr3}.

We then apply a Gaussian mixture model to find the best local shifts of the positions in GDR2 to the positions in GDR3 and the uncertainty inflation term $\alpha$.
After the GDR2 positions have been locally calibrated against GDR3, we calculate proper motions using the time deltas between the central epochs of each measurement in GDR2 and in GDR3. We then once more fit a Gaussian mixture model between the GDR3-GDR2 position derived proper motions and the Gaia GDR3 proper motion reference frame.

After the local calibration of the GDR3-GDR2 position-derived proper motions, we again address a magnitude dependence in the uncertainties of bright stars. We apply the same mixture model globally, fitting only for an uncertainty inflation faction $\alpha$ for each of the magnitude bins $(0,3], (3,5)], (5,7], (7,13],$ and $(13,16)$.

Our corrections applied to GDR3--GDR2 scaled position difference are visualized in Figure \ref{fig:dr32-cal}. As with the GDR2 proper motions, there are several strong systematic patterns visible resulting from the Gaia GDR2 scan law. We note that the uncertainty inflation derived for these quantities is on the whole close to approximately 1.5, indicating fairly well-calibrated position uncertainties.

Figure \ref{fig:zscores-dr32} plots the Z-scores between the calibrated GDR3--GDR2 scaled position differences and the proper motions from GDR3. Here, we find that the calibrated scaled position differences are well matched to the GDR3 proper motions across magnitudes, with the exception that stars fainter than approximately 14th magnitude appear to have their uncertainties somewhat overestimated.

\subsection{Gaia Astrometric Excess Noise}\label{sec:excess-noise}

In order to determine stellar parallaxes, positions, and proper motions, the Gaia team fits an astrometric model to the time series of individual measurements. These are referred to as either 5-parameter or 6-parameter solutions. Each of these individual time series measurements is associated with its own formal uncertainty. During the fit, the uncertainties on the measurements are inflated in quadrature by the \emph{astrometric excess noise} such that the final reduced $\chi^2$ is set to unity. The amount of astrometric excess noise is reported for each solution through the \verb|astrometric_excess_noise| column and,  indirectly, through the \verb|ruwe| column. 
If a companion perturbs a star's straight-line motion through space, as viewed from the Earth, the 5 or 6-parameter model will fit less well and require an increased astrometric excess. Alternatively, a low astrometric excess noise can indicate the \emph{lack} of any significant perturbations over the Gaia DR3 baseline. 

Two barriers stand in the way of modeling the astrometric excess noise directly. The first is that GDR3's formal uncertainties vary from measurement to measurement and are not reported directly. The second is that the amount of excess noise expected for single stars without any companions is non-zero. In order to use astrometric excess noise as a reliable constraint, both of these expected noise levels must be determined.

What follows is a discussion of our implementation of the methodology from \citet{pmex1}, and in particular, the unbiased estimator of variance (UEVA). We will refer to this in general as the constraint coming from Gaia's astrometric excess noise, though not specifically the \verb|astrometric_excess_noise| column reported in GDR3. By default, Octofitter is configured to derived the astrometric excess noise from the RUWE (renormalized unit weight error) for the reasons described in \citet{pmex1}.

Following the method outlined by \citet{pmex1}, we independently calculate estimates of the Gaia formal along-scan uncertainty, $\sigma_{AL}$, as a function of g-band magnitude and bp-rp colour; the excess attitude uncertainty,  $\sigma_{att}$, as a function of R.A.\ and Dec.; and excess calibration uncertainty, $\sigma_{cal}$, as a function of g-band magnitude and bp-rp color. These must be determined for both 5 and 6 parameter solutions. 

We follow the steps outlined by \citet{pmex1}, in Section 3 of their paper, to calculate these median uncertainty maps. Extending beyond the values estimated in \citet{pmex1}, we also estimate uncertainty maps for each of the noise parameters to capture the dispersion in each bin. We refer to these as $\sigma_{AL,\sigma}$, $\sigma_{att,\sigma}$ and $\sigma_{cal,\sigma}$. To ensure that we have reasonably well defined uncertainties we only consider magnitude/colour bins that contain greater than 10 stars. For $\sigma_{AL}$ and $\sigma_{att}$ we estimate the uncertainties as 1.4826 times the median absolute deviation (MAD) in each bin, where 1.4826 is the conversion factor from MAD to standard deviation for a  normal distribution. We use this estimate for the uncertainty because the median absolute deviation is more robust against outliers than the standard deviation. 
For $\sigma_{cal}$, we conservatively estimate its uncertainty map using the 16th percentile of the distribution in each bin. We use this estimate because $\sigma_{cal}$ is contaminated by astrometrically accelerating stars and shows a significant tail on the right end of its distribution.

\begin{figure}
    \centering
    \includegraphics[width=\linewidth]{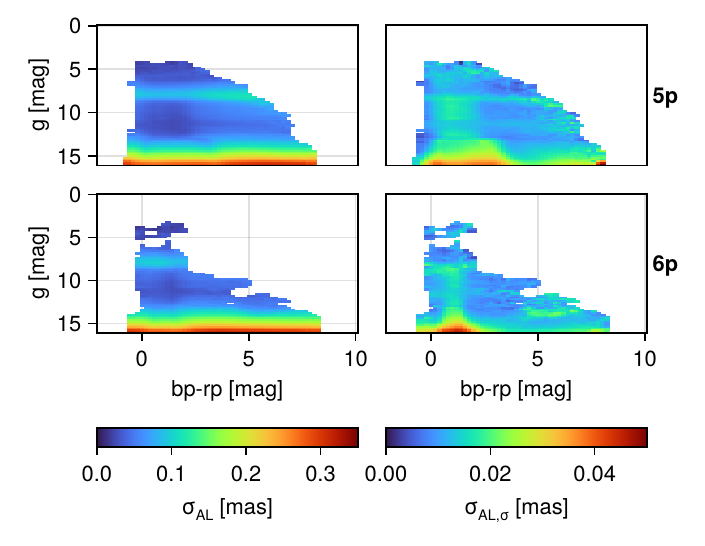}
    \caption{The along scan uncertainty ($\sigma_{AL}$) we estimate for GDR3, following \citet{pmex1}. We also estimate the uncertainties in these noise values.}
    \label{fig:sigma-al}
\end{figure}

\begin{figure}
    \centering
    \includegraphics[width=\linewidth]{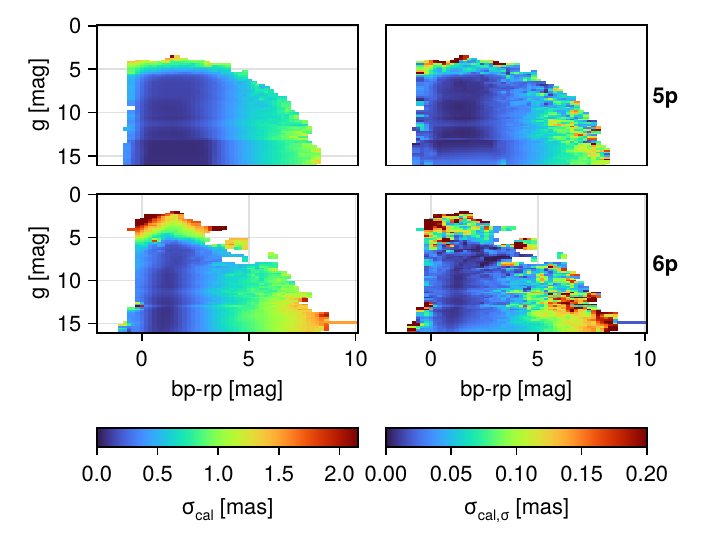}
    \caption{The calibration uncertainty ($\sigma_{cal}$) we estimate for GDR3, following \citet{pmex1}. We also estimate the uncertainties in these noise values.}
    \label{fig:sigma-cal}
\end{figure}

\begin{figure}
    \centering
    \includegraphics[width=\linewidth]{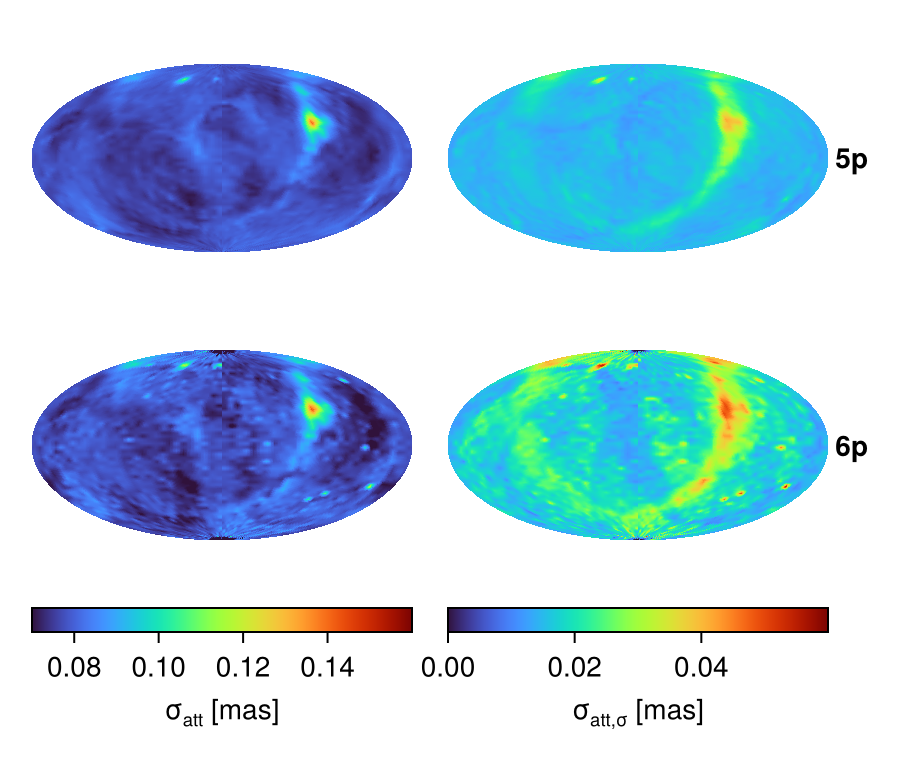}
    \caption{The attitude uncertainty ($\sigma_{att}$) we estimate for GDR3, following \citet{pmex1}. We also estimate the uncertainties in these noise values.}
    \label{fig:sigma-att}
\end{figure}

Our resulting uncertainty maps are presented in Figures \ref{fig:sigma-al}, \ref{fig:sigma-cal}, and \ref{fig:sigma-att}.

To simplify the usage of these maps, we interpolate them for each star in our composite catalog.

\subsection{Example Velocity Curves}

\begin{figure}[ht]
    \begin{adjustwidth}{-1.0cm}{-0.3cm} 
        \includegraphics[width=\linewidth]{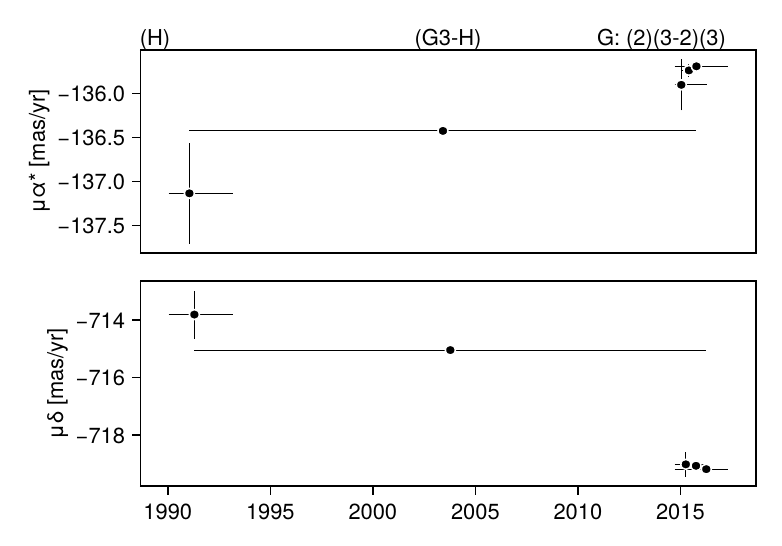}
    \end{adjustwidth}
    \caption{Example astrometric velocity points for Gliese 229 A showing how the long-term trend between Gaia and Hipparcos is resolved into finer time sampling with the addition of the DR2 and DR3-DR2 points. The horizontal lines indicate the time baseline of each measurement, with the centrally propagated epochs marked by points. Labels along the top axis indicate the source of each point.
    This smooth long term motion is attributable to the Gliese 229 Ba/Bb brown dwarf companion on a wide orbit \citep{gl229_disovery, gl229_xuan}.}
    \label{fig:velplot-gl229}
\end{figure}

We now illustrate the results of our calibration by examining the velocity curves of three systems.
The astrometric velocities of Gliese 229 A are displayed in Figure \ref{fig:velplot-gl229}. The added points derived from DR2 and the DR3--DR2 scaled position difference follow the same smooth trend that is visible from the Hipparcos, Gaia DR3--Hipparcos, and DR3 points from the HGCA \citep{hgca_dr3}. This smooth long term motion is attributable to the Gliese 229 Ba/Bb brown dwarf companion on a wide orbit \citep{gl229_disovery, gl229_xuan}.

\begin{figure}[ht]
    \begin{adjustwidth}{-1.0cm}{-0.3cm} 
        \includegraphics[width=\linewidth]{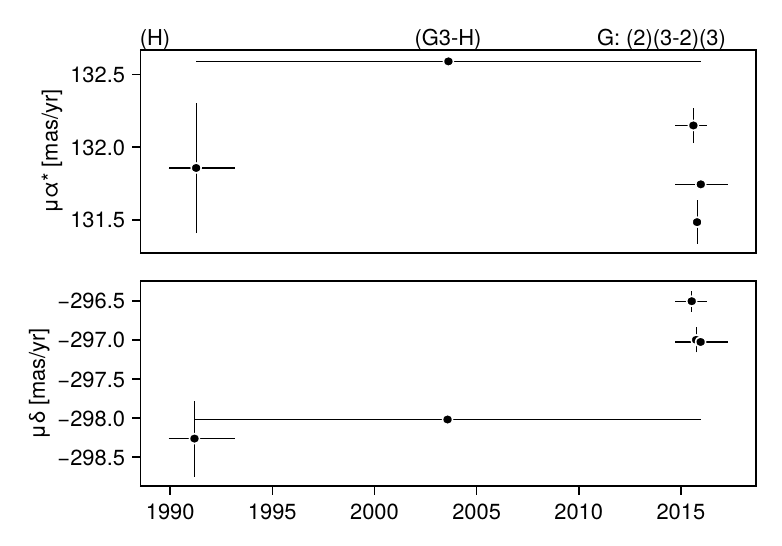}
    \end{adjustwidth}
    \caption{Astrometric velocity points for 14 Her showing both a long-term Gaia-Hipparcos trend attributable to 14 Her c \citep{14herc, 14herc_multi} with an approximately 17 yr orbital period; and significant short term variation, attributable to 14 Her b \cite{14herb}  with a 4.8 yr orbital period (without the shorter period planet, the Gaia-derived points would look more similar to those for Gliese 229 B in Figure \ref{fig:velplot-gl229}. The horizontal lines indicate the time baseline of each measurement, with the centrally propagated epochs marked by points. Labels along the top axis indicate the source of each point.}
    \label{fig:velplot-14her}
\end{figure}

The astrometric velocities of 14 Her displayed in Figure \ref{fig:velplot-14her} are quite different. They appear to contain both a long-term trend, which we attribute to the outer companion 14 Her c \citep{14herc, 14herc_multi}, and short term variation within the Gaia epochs which we attribute to 14 Her b \citep{14herb}, which has a period closer to the Gaia DR3 baseline. This variation would not be detectable looking at the HGCA points alone.

\begin{figure}
    \vspace{1em}
    \begin{adjustwidth}{-1.0cm}{-0.3cm} 
        \includegraphics[width=\linewidth]{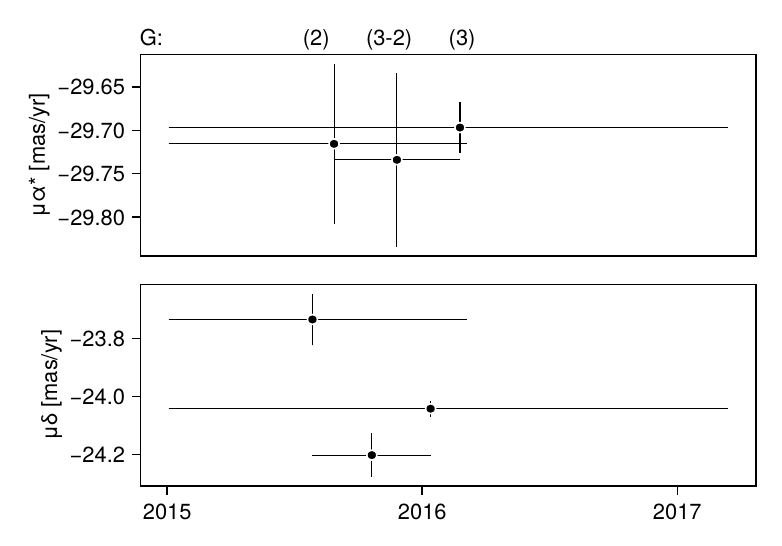}
    \end{adjustwidth}
    \caption{Astrometric velocity points for PDS 70, a source with no Hipparcos data available. By comparing the calibrated GDR2 proper motions, the GDR3 - GDR2 scaled position difference, and the GDR3 proper motions, it would be possible to constrain the astrometric reflex motion of the star, which we leave for a future work.}
    \label{fig:velplot-pds70}
\end{figure}

The astrometric velocities of PDS 70 are displayed in Figure \ref{fig:velplot-pds70}. This system was not cataloged by Hipparcos, and so no HGCA data points are available. The derived astrometric velocities indicate significant reflex motion on timescales of approximately one year in the Dec.\ axis. We leave interpretation of these particular signals to a future work.

\subsection{RV Excess Noise}\label{sec:rv-excess}

For stars brighter than 12th magnitude, GDR3 extracts the radial velocity of the star separately for each recorded spectrum, and then averages the resulting RV. The uncertainty on this value is derived from the sample variance. Therefore, a star with inflated RV variability, e.g.\ due to a massive, close-in companion, results in a higher reported uncertainty on the system's average RV. 

This RV excess noise was used by \citet{paired} to detect a large number of stellar binaries. We have the opposite goal: to rule out close in stellar binaries by a \emph{lack} of significant RV variability. In our implementation, we directly adopt the per-transit RV uncertainties ($\sigma_{\rm RV}$) and our uncertainty error in those uncertainties  ($\sigma_{\rm RV,err}$) estimated in the \citet{paired} catalog. 


\subsection{Catalog Output}\label{sec:catalog-output}

The catalog is provided in Apache Arrow (feather) format which can be accessed through various scripting languages. The catalog is hosted at \href{http://dx.doi.org/10.11570/26.0002}{http://dx.doi.org/10.11570/26.0002}. The columns are described in Table \ref{tab:catalog}.
We offer the following notes on its usage.

First, as a quality check, we recommend skepticism for stars where the \verb|rho_dr2_dr3_cat| value approaches 1. This indicates that the GDR2 and GDR3 have very similar numbers of matched observations, and potentially even the exact same observations. In this case, changes between the GDR2, GDR3-GDR2, and GDR3 proper motions are not well defined. For these stars, we recommend fitting without the GDR2 and GDR32 epochs (i.e. using the HGCA PMA, IAD, UEVA, and RV variability only).

In addition, we caution that the \verb|pmra_hip|, \verb|pmdec_hip|, \verb|pmra_hg|, \verb|pmdec_hg| values retain corrections for non-linear motion between their measurement epochs e.g.\ due to secular acceleration (the apparent non-linear motion caused by curvature in the R.A. / Dec. coordinate system). Our preference is to include the non-linearity correction in the modeling code itself, so we have not applied these same corrections to the other proper motion fields we derive. A potential user comparing the values directly and not using Octofitter should apply their own correction to the DR2 and DR3 derived quantities. Those wishing to remove the corrections already applied in the HGCA should subtract the fields \verb|nonlinear_dpmra| and \verb|nonlinear_dpmdec| from the \verb|pmra_hg| and \verb|pmdec_hg| fields, and subtract double the \verb|nonlinear_dpmra| and \verb|nonlinear_dpmdec| values from the \verb|pmra_hip| and \verb|pmdec_hip|.

\startlongtable
\begin{deluxetable*}{lclc}
\tabletypesize{\scriptsize}
\tablewidth{0pt}
\tablecaption{Description of Catalog Contents\label{tab:catalog}}
\tablehead{
    \colhead{Parameter Name} &
    \colhead{Units} &
    \colhead{Description} &
    \colhead{Ref.}
    }
\startdata
\multicolumn{4}{c}{\textbf{Identifiers}} \\
\texttt{gaia\_source\_id} & & {\it Gaia} DR3 source identification number & \\
\texttt{source\_id\_dr2} & & {\it Gaia} DR2 source identification number & \\
\texttt{hip\_id} & & {\it Hipparcos} source identification number (if available) & \\
\hline
\multicolumn{4}{c}{\textbf{Astrometric Parameters (DR3 Catalog Epoch, J2016)}} \\
\texttt{ra} & degrees & Right ascension at DR3 central epoch & \\
\texttt{dec} & degrees & Declination at DR3 central epoch & \\
\texttt{ra\_error} & mas & Right ascension uncertainty at central epoch & \\
\texttt{dec\_error} & mas & Declination uncertainty at central epoch & \\
\texttt{parallax} & mas & {\it Gaia} DR3 parallax \textbf{[from Gaia DR3]} & \\
\texttt{radial\_velocity} & km\,s$^{-1}$ & Radial velocity \textbf{[from Gaia DR3]} & \\
\hline
\multicolumn{4}{c}{\textbf{Proper Motions}} \\
\texttt{pmra\_hip} & mas\,yr$^{-1}$ & {\it Hipparcos} proper motion in RA \textbf{[from HGCA]} & \\
\texttt{pmra\_hg} & mas\,yr$^{-1}$ & {\it Hipparcos}-{\it Gaia} proper motion in RA \textbf{[from HGCA]} & \\
\texttt{pmra\_dr2} & mas\,yr$^{-1}$ & Cross-calibrated {\it Gaia} DR2 proper motion in RA & \S\ref{sec:cal-dr2-dr3} \\
\texttt{pmra\_dr32} & mas\,yr$^{-1}$ & Proper motion from DR3-DR2 position difference & \S\ref{sec:cal-dr32} \\
\texttt{pmra\_dr3} & mas\,yr$^{-1}$ & {\it Gaia} DR3 proper motion in RA \textbf{[from Gaia DR3]} & \\
\texttt{pmra\_hip\_error} & mas\,yr$^{-1}$ & Uncertainty in \texttt{pmra\_hip} \textbf{[from HGCA]} & \\
\texttt{pmra\_hg\_error} & mas\,yr$^{-1}$ & Uncertainty in \texttt{pmra\_hg} \textbf{[from HGCA]} & \\
\texttt{pmra\_dr2\_error} & mas\,yr$^{-1}$ & Calibrated uncertainty in \texttt{pmra\_dr2} & \S\ref{sec:cal-dr2-dr3} \\
\texttt{pmra\_dr32\_error} & mas\,yr$^{-1}$ & Calibrated uncertainty in \texttt{pmra\_dr32} & \S\ref{sec:cal-dr32} \\
\texttt{pmra\_dr3\_error} & mas\,yr$^{-1}$ & Calibrated uncertainty in \texttt{pmra\_dr3} (inflated by 1.37) & \S\ref{sec:cal-dr2-dr3} \\
\texttt{pmdec\_hip} & mas\,yr$^{-1}$ & {\it Hipparcos} proper motion in Dec \textbf{[from HGCA]} & \\
\texttt{pmdec\_hg} & mas\,yr$^{-1}$ & {\it Hipparcos}-{\it Gaia} proper motion in Dec \textbf{[from HGCA]} & \\
\texttt{pmdec\_dr2} & mas\,yr$^{-1}$ & Cross-calibrated {\it Gaia} DR2 proper motion in Dec & \S\ref{sec:cal-dr2-dr3} \\
\texttt{pmdec\_dr32} & mas\,yr$^{-1}$ & Proper motion from DR3-DR2 position difference & \S\ref{sec:cal-dr32} \\
\texttt{pmdec\_dr3} & mas\,yr$^{-1}$ & {\it Gaia} DR3 proper motion in Dec \textbf{[from Gaia DR3]} & \\
\texttt{pmdec\_hip\_error} & mas\,yr$^{-1}$ & Uncertainty in \texttt{pmdec\_hip} \textbf{[from HGCA]} & \\
\texttt{pmdec\_hg\_error} & mas\,yr$^{-1}$ & Uncertainty in \texttt{pmdec\_hg} \textbf{[from HGCA]} & \\
\texttt{pmdec\_dr2\_error} & mas\,yr$^{-1}$ & Calibrated uncertainty in \texttt{pmdec\_dr2} & \S\ref{sec:cal-dr2-dr3} \\
\texttt{pmdec\_dr32\_error} & mas\,yr$^{-1}$ & Calibrated uncertainty in \texttt{pmdec\_dr32} & \S\ref{sec:cal-dr32} \\
\texttt{pmdec\_dr3\_error} & mas\,yr$^{-1}$ & Calibrated uncertainty in \texttt{pmdec\_dr3} (inflated by 1.37) & \S\ref{sec:cal-dr2-dr3} \\
\hline
\multicolumn{4}{c}{\textbf{Central Measurement Epochs}} \\
\texttt{epoch\_ra\_hip} & year & Central epoch of {\it Hipparcos} RA \textbf{[from HGCA]} & \\
\texttt{epoch\_ra\_hg} & year & Mean epoch of {\it Hipparcos}-{\it Gaia} RA & \S\ref{sec:hgca} \\
\texttt{epoch\_ra\_dr2} & year & Central epoch of {\it Gaia} DR2 RA & \S\ref{sec:cal-dr2-dr3} \\
\texttt{epoch\_ra\_dr32} & year & Mean epoch of DR3-DR2 RA & \S\ref{sec:cal-dr32} \\
\texttt{epoch\_ra\_dr3} & year & Central epoch of {\it Gaia} DR3 RA & \S\ref{sec:cal-dr2-dr3} \\
\texttt{epoch\_dec\_hip} & year & Central epoch of {\it Hipparcos} Dec \textbf{[from HGCA]} & \\
\texttt{epoch\_dec\_hg} & year & Mean epoch of {\it Hipparcos}-{\it Gaia} Dec & \S\ref{sec:hgca} \\
\texttt{epoch\_dec\_dr2} & year & Central epoch of {\it Gaia} DR2 Dec & \S\ref{sec:cal-dr2-dr3} \\
\texttt{epoch\_dec\_dr32} & year & Mean epoch of DR3-DR2 Dec & \S\ref{sec:cal-dr32} \\
\texttt{epoch\_dec\_dr3} & year & Central epoch of {\it Gaia} DR3 Dec & \S\ref{sec:cal-dr2-dr3} \\
\hline
\multicolumn{4}{c}{\textbf{Correlations}} \\
\texttt{pmra\_pmdec\_hip} & & Correlation between Hip RA/Dec PMs \textbf{[from HGCA]} & \\
\texttt{pmra\_pmdec\_hg} & & Correlation between HG RA/Dec PMs \textbf{[from HGCA]} & \\
\texttt{pmra\_pmdec\_dr2} & & Correlation between DR2 RA/Dec PMs & \S\ref{sec:cal-dr2-dr3} \\
\texttt{pmra\_pmdec\_dr32} & & Correlation between DR32 RA/Dec PMs (set to 0) & \S\ref{sec:cal-dr32} \\
\texttt{pmra\_pmdec\_dr3} & & Correlation between DR3 RA/Dec PMs & \S\ref{sec:cal-dr2-dr3} \\
\hline
\multicolumn{4}{c}{\textbf{Excess Noise and Calibration Parameters}} \\
\texttt{ruwe\_dr3} & & Renormalized unit weight error \textbf{[from Gaia DR3]} & \S\ref{sec:ueva} \\
\texttt{astrometric\_excess\_noise\_dr3} & mas & \textbf{[from Gaia DR3]} & \S\ref{sec:ueva} \\
\texttt{sig\_AL} & mas & Along-scan uncertainty floor & \S\ref{sec:excess-noise} \\
\texttt{sig\_AL\_sigma} & mas & Uncertainty in \texttt{sig\_AL} & \S\ref{sec:excess-noise} \\
\texttt{sig\_att\_radec} & mas & Attitude uncertainty & \S\ref{sec:excess-noise} \\
\texttt{sig\_att\_radec\_sigma} & mas & Uncertainty in \texttt{sig\_att\_radec} & \S\ref{sec:excess-noise} \\
\texttt{sig\_cal} & mas & Calibration uncertainty & \S\ref{sec:excess-noise} \\
\texttt{sig\_cal\_sigma} & mas & Uncertainty in \texttt{sig\_cal} & \S\ref{sec:excess-noise} \\
\hline
\multicolumn{4}{c}{\textbf{Radial Velocity}} \\
\texttt{rv\_renormalised\_gof\_dr3} & & \textbf{[from \textit{paired}]} & \\
\texttt{rv\_chisq\_pvalue\_dr3} & & \textbf{[from \textit{paired}]} & \\
\texttt{rv\_amplitude\_robust\_dr3} & & \textbf{[from \textit{paired}]} & \\
\texttt{rv\_ln\_uncert\_dr3} & $\ln(\mathrm{km/s})$ & \textbf{[from \textit{paired}]} & \S\ref{sec:rv-excess} \\
\texttt{rv\_ln\_uncert\_err\_dr3} & $\ln(\mathrm{km/s})$ & \textbf{[from \textit{paired}]} & \S\ref{sec:rv-excess} \\
\texttt{rv\_semiamp\_p5\_dr3} & km/s & \textbf{[from \textit{paired}]} & \\
\texttt{rv\_semiamp\_p16\_dr3} & km/s & \textbf{[from \textit{paired}]} & \\
\texttt{rv\_semiamp\_p50\_dr3} & km/s & \textbf{[from \textit{paired}]} & \\
\texttt{rv\_semiamp\_p84\_dr3} & km/s & \textbf{[from \textit{paired}]} & \\
\texttt{rv\_semiamp\_p95\_dr3} & km/s & \textbf{[from \textit{paired}]} & \\
\texttt{radial\_velocity\_error} & km/s & Uncertainty in Gaia's radial velocity measurement \textbf{[from Gaia DR3]} & \\
\texttt{rv\_nb\_transits} & & Number of samples used to compute radial velocity \textbf{[from Gaia DR3]} & \\
\hline
\multicolumn{4}{c}{\textbf{Misc.}} \\
\texttt{astrometric\_matched\_transits\_dr3} & & \textbf{[from Gaia DR3]} & \S\ref{sec:missed-epochs} \\
\texttt{astrometric\_params\_solved\_dr3} & & \textbf{[from Gaia DR3]} & \\
\texttt{bp\_rp\_dr3} & mag & BP-RP color \textbf{[from Gaia DR3]} & \\
\texttt{phot\_g\_mean\_mag\_dr3} & mag & G-band magnitude \textbf{[from Gaia DR3]} & \\
\texttt{nonlinear\_dpmra} & mas\,yr$^{-1}$ & Nonlinear motion correction \textbf{[from HGCA]} & \S\ref{sec:catalog-output} \\
\texttt{nonlinear\_dpmdec} & mas\,yr$^{-1}$ & Nonlinear motion correction \textbf{[from HGCA]} & \S\ref{sec:catalog-output} \\
\texttt{nss\_acceleration\_astro\_nss\_solution\_type\_dr3} & & \textbf{[from Gaia DR3]} & \\
\texttt{nss\_non\_linear\_spectro\_nss\_solution\_type\_dr3} & & \textbf{[from Gaia DR3]} & \\
\texttt{astrometric\_n\_obs\_al\_dr3} & & \textbf{[from Gaia DR3]} &  \\
\texttt{astrometric\_n\_obs\_ac\_dr3} & & \textbf{[from Gaia DR3]} &  \\
\texttt{astrometric\_n\_good\_obs\_al\_dr3} & & \textbf{[from Gaia DR3]} &  \\
\texttt{astrometric\_n\_bad\_obs\_al\_dr3} & & \textbf{[from Gaia DR3]} & \\
\texttt{astrometric\_gof\_al\_dr3} & & \textbf{[from Gaia DR3]} &  \\
\texttt{astrometric\_chi2\_al\_dr3} & & \textbf{[from Gaia DR3]} & \\
\texttt{astrometric\_excess\_noise\_dr3} & & \textbf{[from Gaia DR3]} &  \\
\texttt{astrometric\_params\_solved\_dr3} & & \textbf{[from Gaia DR3]} & \\
\texttt{new\_matched\_transits\_dr3} & & \textbf{[from Gaia DR3]} &  \\
\texttt{matched\_transits\_removed\_dr3} & & \textbf{[from Gaia DR3]} & \\
\texttt{ipd\_gof\_harmonic\_amplitude\_dr3} & & \textbf{[from Gaia DR3]} & \\
\texttt{ipd\_gof\_harmonic\_phase\_dr3} & & \textbf{[from Gaia DR3]} & \\
\enddata
\tablecomments{Columns marked with \textbf{[from HGCA]} are taken directly from the HGCA DR3 catalog \citep{hgca_dr3}. Columns marked with \textbf{[from Gaia DR3]} are taken directly from Gaia DR3 \citep{gaia_dr3}. Columns marked with \textbf{[from paired]} are taken directly from the catalog published in \citet{paired}. All other columns are derived in this work through cross-calibration. Note: the calibrated GDR2 proper motions and GDR3-GRD2 scaled position differences do not have any correction applied for secular acceleration, while the fields from the HGCA do.}
\end{deluxetable*}

\section{Modeling}\label{sec:modelling}

We implement models of the catalog quantities described above in Octofitter\footnote{\href{https://sefffal.github.io/Octofitter.jl/}{https://sefffal.github.io/Octofitter.jl/}} \citep{octofitter}, a Julia \citep{julia_bezanson} package for Bayesian modeling of exoplanet data.

The total log-likelihood is the sum of four components: 
\begin{equation}
\ln \mathcal{L} = \ln \mathcal{L}_{\rm UEVA}  + \ln \mathcal{L}_{\rm PM} + \ln \mathcal{L}_{\rm IAD} + \ln \mathcal{L}_{\rm RV},
\end{equation}
corresponding to the Gaia astrometric excess noise (unbiased estimator of variance, or UEVA), the proper motion epochs,  the Hipparcos intermediate astrometric data (IAD), and the Gaia radial velocity variability, respectively. Throughout, tildes denote observed catalog quantities and unadorned symbols denote model predictions. We write $\ln \mathcal{N}(\tilde{x} \mid x, \Sigma)$ for the log probability density of a (multivariate) normal distribution with mean $x$ and covariance $\Sigma$, evaluated at $\tilde{x}$; this function notation is standard in the statistical literature.


\subsection{Gaia Measurement Epochs}\label{sec:missed-epochs}

To forward-model the Gaia reductions, we need estimates of the visibility windows and scan angles at which Gaia observed the target. We use the Gaia Observation Forecast Tool (GOST)\footnote{\href{https://gaia.esac.esa.int/gost/index.jsp}{https://gaia.esac.esa.int/gost/index.jsp}}, with Octofitter providing functionality to automatically query GOST for any given target.

In most cases, the true number of matched transits given by the \verb|astrometric_matched_transits| field ($N_{\rm FoV}$)  is less than the number of epochs returned by GOST. That is, some measurements are skipped due to gaps in the Gaia data or are rejected as outliers. 
To address this, we treat the selection of a subset of $N_{\rm FoV}$ epochs as a random variable which is sampled as part of the MCMC algorithm. A discrete uniform prior over subsets of size $N_{\rm AL}$ can be equivalently formulated as follows: first, introduce an auxiliary random variable, called a ``priority,'' for each possible epoch, with a standard Gaussian prior; second, select the epochs and scan angles from GOST that correspond to the top $N_{\rm AL}$ priorities to use for the astrometry modeling. This allows us to marginalize over the uncertainty in which discrete epochs are used for the subsequent simulations and modeling.



\subsection{UEVA Likelihood}\label{sec:ueva}

The UEVA constrains the amount of within-Gaia astrometric scatter, following \citet{pmex1}. Following that work, we parameterize the Gaia per-observation uncertainty (which we assume is equal for every observation) with three components drawn from priors informed by our noise calibration maps: an along-scan photon noise floor $\sigma_{\rm AL}$, an attitude uncertainty $\sigma_{\rm att}$, and a calibration uncertainty $\sigma_{\rm cal}$. The formal per-observation uncertainty is $\sigma_{\rm formal} = \sqrt{\sigma_{\rm AL}^2 + \sigma_{\rm att}^2}$.

The UEVA likelihood compares the catalog and model UEVA values using a cube-root transformation to improve normality:
\begin{equation}
\ln \mathcal{L}_{\rm UEVA} = \ln \mathcal{N}\left(\tilde{U}^{1/3} \,\Big|\, U_{\rm model}^{1/3}, \sigma_U\right)
\end{equation}
where $\sigma_U = \sigma_{U,\rm single}/(3 \, \langle U \rangle_{\rm single}^{2/3})$.

The catalog UEVA can be computed from either the astrometric excess noise ($\tilde{U} = \epsilon_{\rm astro}^2 + \sigma_{\rm att}^2 + \sigma_{\rm AL}^2$) or the RUWE. For a single star with no companions, the expected UEVA is:
\begin{equation}
\langle U \rangle_{\rm single} = \frac{N_{\rm AL/FoV}}{N_{\rm AL} - n_{\rm dof}} \left[ (N_{\rm FoV} - n_{\rm dof}) \sigma_{\rm cal}^2 + N_{\rm FoV} \sigma_{\rm AL}^2 \right]
\end{equation}
where $N_{\rm FoV}$ is the number of field-of-view transits and $N_{\rm AL/FoV} = N_{\rm AL} / N_{\rm FoV}$ is the average number of along-scan measurements per transit.

To compute the model-predicted UEVA, we evaluate the chi-squared of the along-scan residuals after fitting a 5-parameter (single-star) astrometric solution to the perturbed sky path, using the drawn subset of observing epochs described in Section~\ref{sec:missed-epochs}:
\begin{equation}
U_{\rm model} = \left( \frac{\chi^2_{\rm model} \cdot \sigma_{\rm formal}^2}{N_{\rm AL/FoV} \cdot N_{\rm FoV} - n_{\rm dof}} + \langle U \rangle_{\rm single} \right)^{1/3}.
\end{equation}

\subsection{Proper Motion Likelihood}\label{sec:pm-likelihood}

As with previous codes like \texttt{orvara} \citep{orvara}, we forward-model how a perturbation from a companion would affect the astrometric fits from Hipparcos, GDR2, and GDR3. Starting with a set of measurement epochs (taken from the Hipparcos IAD or the Gaia scan law; see Section~\ref{sec:missed-epochs}), we simulate the along-scan perturbations caused by a set of orbital parameters. We then perform linear 5-parameter astrometric fits over the Hipparcos, GDR2, and GDR3 time ranges, yielding derived perturbations to each catalog's positions, proper motions, and parallaxes. We add these values to the model's barycentric values and compare the perturbed proper motions with the calibrated catalog values. We  include parallax in the linear fits but not the outer model likelihood as the differences between catalogs are dominated by systematics \citep{lindegren_parallax_bias}.

This forward-modeling approach yields model-predicted proper motions $\boldsymbol{\mu}_{\rm H}$, $\boldsymbol{\mu}_{\rm HG}$, $\boldsymbol{\mu}_{\rm DR2}$, $\boldsymbol{\mu}_{\rm DR32}$, $\boldsymbol{\mu}_{\rm DR3}$, and a predicted UEVA value $U_{\rm model}$, where each $\boldsymbol{\mu}$ is a two-component vector containing the R.A.\ and Dec.\ proper motions.

The block structure of the covariance matrix allows us to decompose the proper motion likelihood into five terms:
\begin{equation}
\ln \mathcal{L}_{\rm PM} = \ln \mathcal{L}_{\rm H} + \ln \mathcal{L}_{\rm HG} + \ln \mathcal{L}_{\rm DR2,DR3} + \ln \mathcal{L}_{\rm DR32}.
\end{equation}

The Hipparcos proper motion term is:
\begin{equation}
\ln \mathcal{L}_{\rm H} = \ln \mathcal{N}\left(\tilde{\boldsymbol{\mu}}_{\rm H} \,\Big|\, \boldsymbol{\mu}_{\rm H}, \boldsymbol{\Sigma}_{\rm H}\right),
\end{equation}
where $\boldsymbol{\Sigma}_{\rm H}$ is the $2 \times 2$ covariance matrix from the HGCA incorporating the R.A./Dec.\ correlation.

The long-baseline Hipparcos--Gaia proper motion term is:
\begin{equation}
\ln \mathcal{L}_{\rm HG} = \ln \mathcal{N}\left(\tilde{\boldsymbol{\mu}}_{\rm HG} \,\Big|\, \boldsymbol{\mu}_{\rm HG}, \boldsymbol{\Sigma}_{\rm HG}\right),
\end{equation}
where $\boldsymbol{\Sigma}_{\rm HG}$ is again taken from the HGCA. We remove the perspective acceleration corrections applied within the HGCA in favour of modeling this effect within Octofitter.

We now turn to the Gaia-only proper motion quantities. We use results from UEVA modeling (Section~\ref{sec:ueva}) to improve our proper motion constraints at the GDR3 and GDR3-GDR2 epochs. Since we have carefully modeled how the GDR3 astrometric uncertainties were inflated by orbital motion compared to what is expected for a typical single star, we can, by continuing to assume that the uncertainties are equal across measurement epochs, \emph{de-inflate} the reported GDR3 parameter uncertainties by a multiplicative factor:
\begin{equation}\label{eq:deflate}
f_{\rm deflate} = \min\left(1, \sqrt{\langle U \rangle_{\rm single}/\tilde{U}}\right)
\end{equation}

Because DR2 and DR3 share a significant fraction of their observations, these proper motions are correlated. We model them jointly following \citet{thompson_eps_eri}:
\begin{equation}
\ln \mathcal{L}_{\rm DR2,DR3} = \ln \mathcal{N}\left( \begin{pmatrix} \tilde{\boldsymbol{\mu}}_{\rm DR2} \\ \tilde{\boldsymbol{\mu}}_{\rm DR3} \end{pmatrix} \,\Bigg|\, \begin{pmatrix} \boldsymbol{\mu}_{\rm DR2} \\ \boldsymbol{\mu}_{\rm DR3} \end{pmatrix}, \boldsymbol{\Sigma}_{\rm DR2,DR3} \right)
\end{equation}

We define the $4 \times 4$ joint covariance matrix as:
\begin{equation}
\boldsymbol{\Sigma}_{\rm DR2,DR3} = \begin{pmatrix} \boldsymbol{\Sigma}_{\rm DR2} & \mathbf{K} \\ \mathbf{K}^T & \boldsymbol{\Sigma}_{\rm DR3}' \end{pmatrix}
\end{equation}
where $\boldsymbol{\Sigma}_{\rm DR2}$ is the calibrated DR2 proper motion covariance, $\boldsymbol{\Sigma}_{\rm DR3}' = f_{\rm deflate}^2 \boldsymbol{\Sigma}_{\rm DR3}$ is the deflated DR3 covariance (Equation~\ref{eq:deflate}), and the cross-correlation matrix is:
\begin{equation}
\mathbf{K} = \rho_{\rm DR2,DR3} \, \sqrt{\boldsymbol{\Sigma}_{\rm DR2}} \, \sqrt{\boldsymbol{\Sigma}_{\rm DR3}'}
\end{equation}
where $\sqrt{\cdot}$ denotes the matrix square root (not the square root of the matrix elements).


The proper motion derived from the GDR3-GDR2 position difference is modeled as:
\begin{equation}
\ln \mathcal{L}_{\rm DR32} = \ln \mathcal{N}\left(\tilde{\boldsymbol{\mu}}_{\rm DR32} \,\Big|\, \boldsymbol{\mu}_{\rm DR32}, \boldsymbol{\Sigma}_{\rm DR32}'\right)
\end{equation}

The covariance $\boldsymbol{\Sigma}_{\rm DR32}'$ also incorporates the deflation factor (Equation~\ref{eq:deflate}) to account for uncertainty inflation applied by GDR3 that our orbital model now explains. The corrected covariance is:
\begin{equation}
\boldsymbol{\Sigma}_{\rm DR32}' = \boldsymbol{\Sigma}_{\rm DR32} + \Delta\boldsymbol{\Sigma}_{\rm DR32}
\end{equation}
where
\begin{equation}
\begin{split}
\Delta\boldsymbol{\Sigma}_{\rm DR32} &= \mathbf{T} \biggl[ (f_{\rm deflate}^2 - 1) \boldsymbol{\Sigma}_{\rm pos,DR3} \\
&\quad - (f_{\rm deflate} - 1)(\boldsymbol{\Sigma}_{\rm cross} + \boldsymbol{\Sigma}_{\rm cross}^T) \biggr] \mathbf{T}^T
\end{split}
\end{equation}
Here $\boldsymbol{\Sigma}_{\rm pos,DR3}$ and $\boldsymbol{\Sigma}_{\rm cross}$ are the DR3 position covariance and the DR2--DR3 position cross-covariance at their central epochs, and $\mathbf{T} = \mathrm{diag}(1/\Delta t_\alpha, 1/\Delta t_\delta)$ converts from position to proper motion units.

\subsection{Hipparcos IAD Likelihood}

For stars with Hipparcos observations, we optionally include a likelihood term from the intermediate astrometric data (IAD), retrieved from the ``Java Tool'' for the Hipparcos 2 reduction by \citet{hipparcos_2}. We use this data \emph{in addition} to the net proper motion and position constraints from the HGCA \citep{hgca_dr3}.

The HGCA provides net proper motions and GDR3--Hipparcos scaled position differences that are calculated from a linear combination of two Hipparcos reductions \citet{hipparcos_1} and \citet{hipparcos_2} that have then been calibrated to match the GDR3 reference frame (see Section \ref{sec:hgca}).

The IAD (from either reduction) provide epoch astrometry that can be used to resolve motion within the Hipparcos baseline; however, this data is not calibrated against the GDR3 reference frame, and in addition \citet{brandt_iad} notes that ``that a merged set of IAD from the 1997 and 2007 Hipparcos reductions is not possible in an internally consistent manner.''

To avoid inconsistency with the composite HGCA proper motions, we fit only the acceleration information in the IAD by introducing nuisance parameters $(\Delta\alpha_{\rm IAD}, \Delta\delta_{\rm IAD}, \Delta\varpi_{\rm IAD}, \mu_{\alpha*,\rm IAD}, \mu_{\delta,\rm IAD})$ for a local reference frame, analogous to the use of offsets and linear trends in radial velocity modeling. The IAD likelihood is:
\begin{equation}
\ln \mathcal{L}_{\rm IAD} = \sum_{i=1}^{N_{\rm IAD}} \ln \mathcal{N}\left(r_i \,\Big|\, 0, \sqrt{\sigma_i^2 + \sigma_{\rm jitter}^2}\right),
\end{equation}
where $r_i$ is the residual for the $i$-th non-rejected observation, $\sigma_i$ is the adjusted reported uncertainty, and $\sigma_{\rm jitter}$ is a parameter we introduce representing additional astrometric noise. 

In addition, we follow the recommendations of \citet{brandt_iad} by adding 0.140~mas to each residual and 2.25~mas in quadrature to each uncertainty to mitigate known overfitting.


\subsection{Gaia RV Variability Likelihood}

Given proposed orbital parameters, we evaluate the RV curve at Gaia's observing epochs and compute its sample variance. This yields the non-centrality parameter $\lambda = (N_{\rm RV} - 1) s^2_{\rm model} / \sigma_{\rm RV}^2$ described in \citet{paired}.
We then compute the likelihood as the probability density of Gaia's observed variance under a non-central $\chi^2$ distribution:
\begin{equation}
\ln \mathcal{L}_{\rm RV} = \ln f_{\chi^2_{N_{\rm RV}-1}(\lambda)}\left(\tilde{\xi}^2\right)
\end{equation}
where $\tilde{\xi}^2 = (N_{\rm RV} - 1) \tilde{s}^2 / \sigma_{\rm RV}^2$ is the chi-squared statistic corresponding to Gaia's observed sample variance $\tilde{s}^2$. 

As discussed in \ref{sec:missed-epochs}, to select which epochs, we select the GOST epochs at each draw that correspond to the top $N_{\rm RV}$ priorities. This adds the assumption that the epochs used for Gaia's RV measurements are a subset of those used by the astrometry solution.
This differs from the methodology of \citet{paired} who draw the epochs from uniform distributions.


\subsection{Companion Flux}

If the companion is luminous in the visible bands sensed by Gaia or Hipparcos, then its signal (and therefore mass) will be diluted. To illustrate this effect, consider a binary comprised of two stars with exactly equal flux. Their orbits will produce no apparent astrometric signal since their photocenter will remain fixed over time. This is only significant for bright companions that are close enough to the primary that Gaia extracts them as a single sources, instead of two separate sources.

For systems where this consideration is important, Octofitter has the option to account for this flux \cite[as in ][]{pmex1} using any user-provided model. 
In this work, we report only the ``dark mass,'' the mass assuming the companion does not contribute significant flux to Gaia and/or Hipparcos. This is analogous to the use of $m\sin(i)$ in radial velocity modeling, but with a more complex expression.

\subsection{Sampling and Significance}

To compute the significance of a detection, we include both one-companion and zero-companion models within a single MCMC run using a binary indicator variable $z \sim \operatorname{Bernoulli}(0.5)$. A Bayes factor can be computed from $\bar{z}$, the marginal posterior mean of $z$, via $\mathrm{BF} = \bar{z}/(1-\bar{z})$. Following standard guidelines \citep{KassRaftery1995}, a Bayes factor greater than 3 indicates ``strong evidence,'' greater than 10 indicates ``very strong evidence,'' and greater than 100 indicates ``extreme evidence.'' We caution that this metric only considers statistical properties of the data.

We use the non-reversible parallel tempered sampler Pigeons \citep{non_rev_parallel_temp,pigeons} to sample from the posterior distributions, with a slice sampler as the local explorer to handle the discrete parameter $z$ and the discontinuities in the posterior densities caused by the priority random variables. On a compute node with 16 cores, sampling from the combined astrometry posterior takes approximately 30 minutes. The bulk of this compute time results from our inclusion of the Gaia measurement epochs as variables in our MCMC sampler and fixing these brings down the compute time.

The results are full orbital posteriors, not simply separation and mass. When the one-companion model is not strongly favored, the posteriors can still be used to rule out possible companions by calculating, e.g., 95\% upper limits to the mass-separation posterior.

\subsection{Model Parameters}
Table~\ref{tab:priors} summarizes all model parameters and their prior distributions. We parameterize orbits using period $P$ and mass ratio $q = m/M_\star$, with the companion mass $m$ and semi-major axis $a$ derived in post-processing using the host star mass $M_\star$ from the Tess Input Catalog \citep[the TIC;][]{tic} and Kepler's third law. Following \citet{octofitter}, we use the position angle $\theta$ at a reference epoch (taken as the mean of the Gaia observing epochs) to specify the companion's location in an orbit rather than, say, the epoch of periastron passage. As discussed in Section~\ref{sec:missed-epochs}, we introduce one priority variable per candidate epoch from GOST, drawn from a standard normal prior. The majority of parameters in any given system's model are these priorities.

\begin{deluxetable*}{lll}
\tablecaption{Model parameters and prior distributions.\label{tab:priors}}
\tablehead{
\colhead{Parameter} & \colhead{Prior} & \colhead{Description}
}
\startdata
\multicolumn{3}{c}{\textit{Orbital Parameters}} \\
$P$ & $\log\mathcal{U}(1/365.25, 10^4)$~yr & Orbital period \\
$q$ & $\log\mathcal{U}(10^{-5}, 1)$ & Mass ratio $m/M_\star$ \\
$e$ & $\mathcal{U}(0, 0.99)$ & Eccentricity \\
$\omega$ & $\mathcal{U}(0, 2\pi)$ & Argument of periastron \\
$i$ & Sine & Inclination \\
$\Omega$ & $\mathcal{U}(0, 2\pi)$ & Longitude of ascending node \\
$\theta$ & $\mathcal{U}(0, 2\pi)$ & Position angle at reference epoch \\
$z$ & $\mathrm{Bernoulli}(0.5)$ & Companion indicator \\
\hline
\multicolumn{3}{c}{\textit{Astrometric Parameters}} \\
$\varpi$ & $\mathcal{N}^+(\text{catalog}, \text{catalog error})$ & Parallax \\
$\mu_{\alpha*}$ & $\mathcal{U}(\mu_{\alpha*,\rm DR3} \pm 1\text{ mas yr}^{-1})$ & Barycentric PM in R.A. \\
$\mu_\delta$ & $\mathcal{U}(\mu_{\delta,\rm DR3} \pm 1\text{ mas yr}^{-1})$ & Barycentric PM in Dec. \\
\hline
\multicolumn{3}{c}{\textit{Noise Parameters}} \\
$\sigma_{\rm AL}$ & $\mathcal{N}^+(\text{cal.\ map}, \text{cal.\ map error})$ & Along-scan uncertainty \\
$\sigma_{\rm att}$ & $\mathcal{N}^+(\text{cal.\ map}, \text{cal.\ map error})$ & Attitude uncertainty \\
$\sigma_{\rm cal}$ & $\mathcal{N}^+(\text{cal.\ map}, \text{cal.\ map error})$ & Calibration uncertainty \\
$\sigma_{\rm jitter}$ & $\log\mathcal{U}(0.001, 100)$~mas & Hipparcos IAD jitter \\
$\sigma_{\rm RV}$ & $\mathcal{N}^+(\text{catalog}, \text{catalog error})$ & Per-transit RV uncertainty \\
\hline
\multicolumn{3}{c}{\textit{Epoch Selection}} \\
$p_j$ & $\mathcal{N}(0, 1)$ & Priority for epoch $j$ \\
\hline
\multicolumn{3}{c}{\textit{Hipparcos IAD Nuisance Parameters}} \\
$\Delta\alpha_{\rm IAD}$ & $\mathcal{U}(-1000, 1000)$~mas & Position offset in R.A. \\
$\Delta\delta_{\rm IAD}$ & $\mathcal{U}(-1000, 1000)$~mas & Position offset in Dec. \\
$\Delta\varpi_{\rm IAD}$ & $\mathcal{U}(-10, 10)$~mas & Parallax offset \\
$\Delta\mu_{\alpha*,\rm IAD}$ & $\mathcal{U}(-1000, 1000)$~mas~yr$^{-1}$ & PM offset in R.A. \\
$\Delta\mu_{\delta,\rm IAD}$ & $\mathcal{U}(-1000, 1000)$~mas~yr$^{-1}$ & PM offset in Dec. \\
\enddata
\tablecomments{$\mathcal{U}$: uniform; $\log\mathcal{U}$: log-uniform; $\mathcal{N}$: normal; $\mathcal{N}^+$: truncated normal with positive support. The parallax prior is additionally truncated at the lower end to $\max(0, \varpi_{\rm cat} - 10\sigma_{\varpi,\rm cat})$.  ``Cal.\ map'' refers to our calibration maps described in Section~\ref{sec:ueva}.}
\end{deluxetable*}

\section{Results}\label{sec:results}

We now present a series of posterior distributions calculated using the full likelihood described above with data from the HGCA \citep{hgca_dr3}, Hipparcos IAD, our calibrated Gaia DR3 vs.\ DR2 proper motion anomaly, and astrometric excess noise. For convenience, we consider the set of Hipparcos stars that are included in the TIC \citep{tic}, as these have have uniformly estimated host masses available.

\subsection{14 Her}

\begin{figure}
    \centering
    \begin{adjustwidth}{-0.8cm}{-0.8cm} 
    \includegraphics[width=\linewidth]{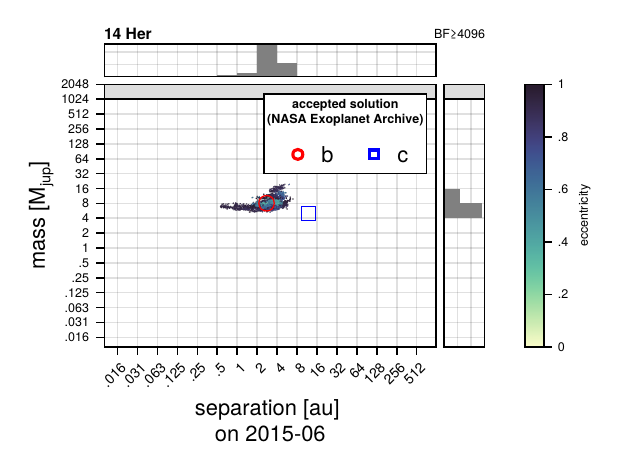}
    \end{adjustwidth}
    \caption{Mass vs.\ separation marginal posterior for HIP 79248 (14 Her). The small points represent draws from the posterior, colored by eccentricity. The horizontal axis is the 3D separation of companion from the star in AU in mid-2015. The larger colored markers present the mass and semi-major axes of the two known planets as retrieved from the accepted entry in the NASA Exoplanet Archive. }
    \label{fig:14-her-1pl}
\end{figure}

We begin by presenting our orbit posterior for 14 Her (aka HIP79248) in Figure \ref{fig:14-her-1pl}.
We visualize the orbit posterior in mass-separation space, where we define separation as the 3D orbital separation in AU between the companion and the star on a particular date. We find that in practice, this quantity is typically better constrained by this data than period or semi-major axis. This may be because these posteriors are often in the regime of fitting partial orbital arcs, which is a similar challenge faced when fitting relative astrometry from direct imaging. The position at the current epoch is typically better constrained than the orbital period.

Figure \ref{fig:14-her-1pl} shows that we can independently confirm planet b using only data from Gaia and Hipparcos. A close-in or wide-separation stellar binary is ruled out without the requirement of additional radial velocity time series or imaging observations. 
This solution is obtained because, referring back to Figure \ref{fig:velplot-14her}, there are a sufficient number of astrometric data points to fit a unique orbit. This is different than previous approaches using only proper motion anomaly or astrometric excess noise, which leave degeneracies between mass and separation.

We also attempted 2- and 3-companion fits to the 14 Her absolute astrometry data, but found only upper limits on the mass of any second companion as a function of separation (e.g., 14 Her c). In Section \ref{sec:14-her-rv}, we will demonstrate a 2-companion fit to absolute astrometry data augmented with time series RVs.

\subsection{Spectroscopic Binary}

An interesting property of combining RV and astrometry constraints is that the two data sources have very different sensitivity scaling behaviors. One effect is that the amplitude of a astrometric perturbation for a given primary and companion falls as the distance to the system increases, while the amplitude of the RV perturbation does not (though our ability to discern the RV perturbation does decrease with decreasing flux). Another effect is that the amplitude of an RV perturbation from a planet with a given mass decreases as the orbital separation grows, meanwhile the amplitude of an astrometric perturbation increases (though our ability to discern the curvature in astrometric perturbation falls suddenly as the period increases beyond our measurement baseline).

\begin{figure}
    \centering
    \begin{adjustwidth}{-0.8cm}{-0.8cm} 
    \includegraphics[width=\linewidth]{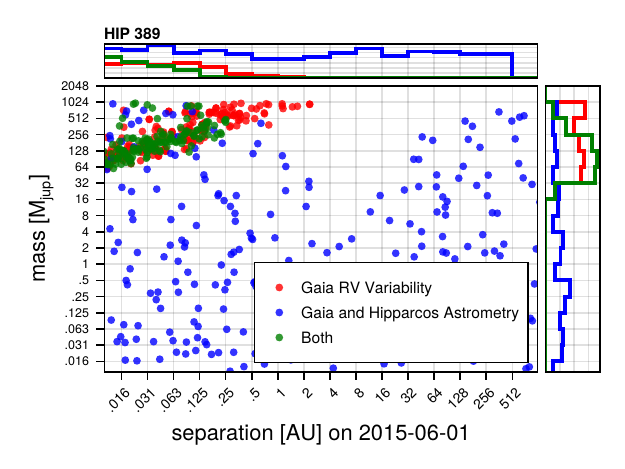}
    \end{adjustwidth}
    \caption{Posterior for the system HIP 389 (aka. Blanco 1-W86), a spectroscopic binary \citep{sb-ref}. This shows how the constraints from Gaia RV variability (via the approach from \citep{paired}) can provide an orthogonal constraint beyond absolute astrometry.}
    \label{fig:sb}
\end{figure}

We demonstrate a fit to a known, distant spectroscopic binary HIP 389 \citep{sb-ref} in Figure \ref{fig:sb}. This system is distant, at over 230pc, so the the absolute astrometry constraints are not sensitive to the close in companion; meanwhile, the constraints from Gaia RV variability do strongly indicate the presence of a companion.

\subsection{Resolved Stellar Binaries}

In order to assess this method visually, we generate posteriors for 75 systems listed in the Sixth Catalog of Orbits of Visual Binary Stars \citep[ORB6;][]{orb6}. We include Hipparcos systems that are in the TESS Input Catalog \citep[TIC;][]{tic} with ORB6 orbit grades of 3 or better and orbital periods less than 50 years. For the primary mass, we assumed the best value from the TESS input catalog.

\begin{figure*}
    \centering
    \includegraphics[width=\linewidth]{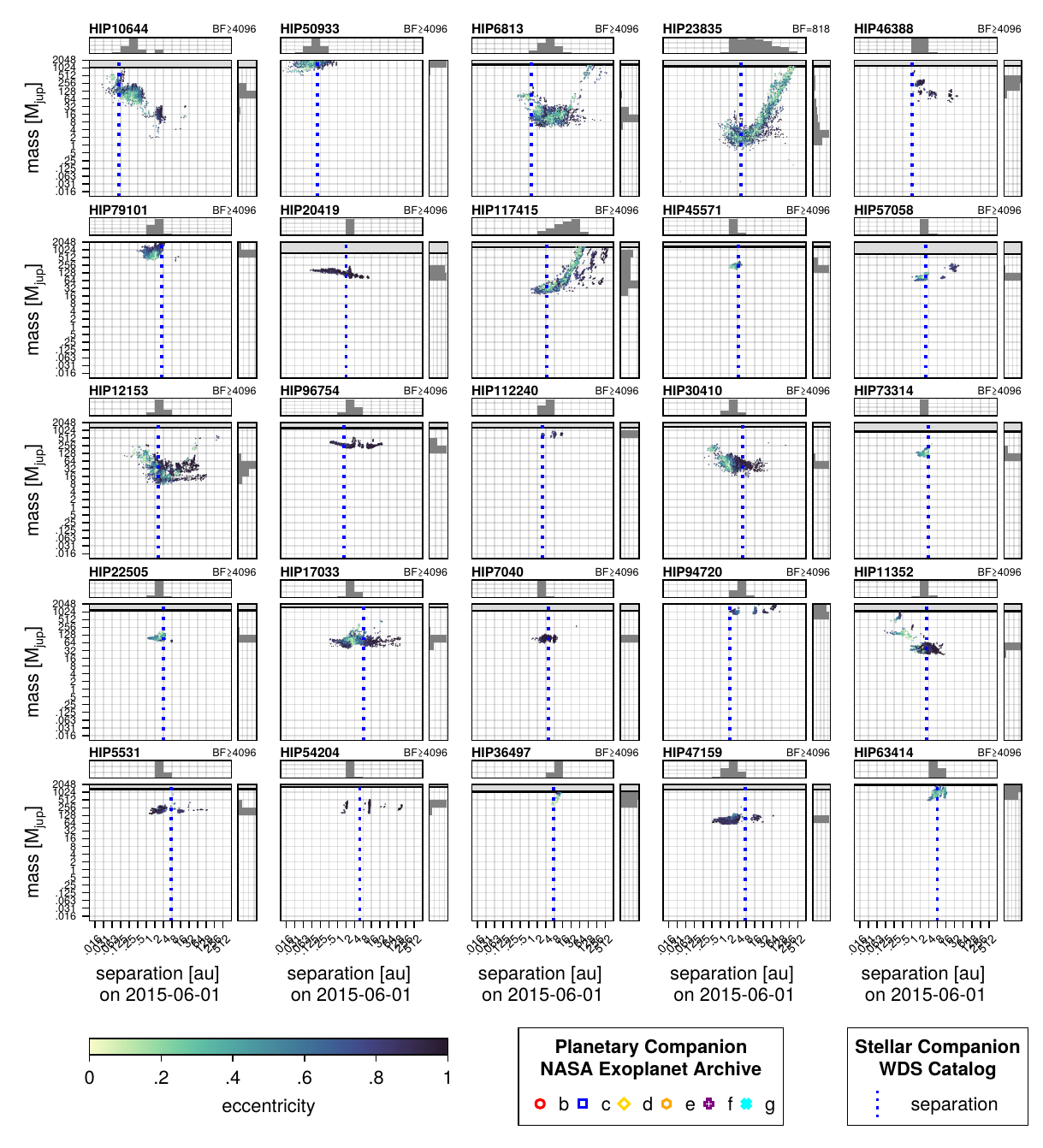}
\caption{Marginal mass-separation posteriors from absolute astrometry for 25 stellar binaries with high quality orbits published in the Washington Double Star (WDS) catalog.  The horizontal gray line indicates the host mass, which is the upper limit of the companion mass prior we adopted. The 3D separation of the companion from the star as calculated from the WDS is indicated as dashed blue lines for each solution. Companion masses are not reported in this catalog, so we indicate the separation only. Our posteriors calculate the ``dark-mass'' of companions assuming they do not significantly contaminate the Gaia photocenter calculations. This assumption breaks down for near equal-brightness binaries, hence the low mass estimates in some cases.}
    \label{fig:wds-stars}
\end{figure*}

We plot the inferred companion mass (assuming a dark companion) against the 3D orbital separation in mid-2015 in Figure \ref{fig:wds-stars}. 
We find that all companions are recovered at high significance and broadly consistent separations. In a few instances such has HIP 6813 and HIP 23835, the posterior prefers a low ``dark mass.'' This is a result of the secondary contributing significant flux, reducing the amplitude of the photocenter shift.

This flux-mass degeneracy  could be addressed through the addition of external information, which we leave for future work. For this paper, we clarify that the mass reported is the ``dark mass'' assuming the flux of the secondary is negligible. This is analogous to the mass-inclination degeneracy in radial velocity measurements.

\subsection{Known Exoplanetary Systems}

As in the previous section, we also test this method against a large set of systems listed on the NASA Exoplanet archive. Our criteria to include a system in this test were that the system was within 40pc, included in the Hipparcos and TIC \citep{tic}, and contained at least one confirmed, non-controversial planet with an orbital period greater than 365 days. We also excluded any stars where the approximate correlation between the GDR2 and GDR3 reductions (calculated in  our \verb|rho_dr2_dr3_cat| column) was greater than 0.999---that is, we exclude systems where the data used in DR2 and DR3 is essentially identical. This leaves 120 unique systems. 
We used a fixed value for the host star's mass as given by the TESS Input Catalog.

The posteriors for all systems are presented in Figures \ref{fig:nea-stars}, \ref{fig:nea-stars-2}, \ref{fig:nea-stars-3}, \ref{fig:nea-stars-4}, and \ref{fig:nea-stars-5}.

We plot the semi-major axis of each known planet using different colored markers for each known planet. We include all solutions from the archive as separate points. Upwards-pointing arrows indicating the value is an $m \sin{i}$ from a radial velocity solution and a plain open marker indicates a true dynamical mass measurement. We do caution that these dynamical mass measurements were usually determined in the literature by fitting Gaia and Hipparcos data e.g.\ via joint fits between time series RVs and the HGCA, so they aren't truly independent measurements. Here, the test is that our method using Gaia and Hipparcos \emph{alone} recovers posteriors consistent with those derived from Gaia and Hipparcos in conjunction with external RV data.

Out of the 120 systems considered, we recover 94 systems where evidence supports a companion with Bayes factors greater than 3 (91 detections with Bayes factors > 10), and 22 ambiguous or non-detections with Bayes factors less than 3. The posteriors for these latter 22 systems remain consistent with the published solutions; i.e., the posteriors do not incorrectly rule out confirmed planets.

In three remaining cases, we instead recover outer brown dwarf and stellar companions. These are Gliese 229 BaBb \citep{gl229_disovery,gl229_xuan}, 
HD 196885 A (20399+1115), 
and HD 7449 B \citep{Feng2022}.
In one case, 55 Cancri, we do not detect planet d, in an apparent false-negative; however, this signal has been called into question in \citet{Harrell_2025}.
We finally note that the listed planetary mass for HD 141937 b, which disagrees with our posterior in Figure \ref{fig:nea-stars-5}, has been debated in the literature. The most relevant work is \citet{kiefer_2020}, which prefers a brown dwarf solution using the PMA and Gaia astrometric excess noise constraints \citep{pmex1} which informed this work, and \citet{Piccinini_arxiv}, which supports a planetary solution but does not rule out that the object is a brown dwarf.

\begin{figure*}
    \centering
    \includegraphics[width=\linewidth]{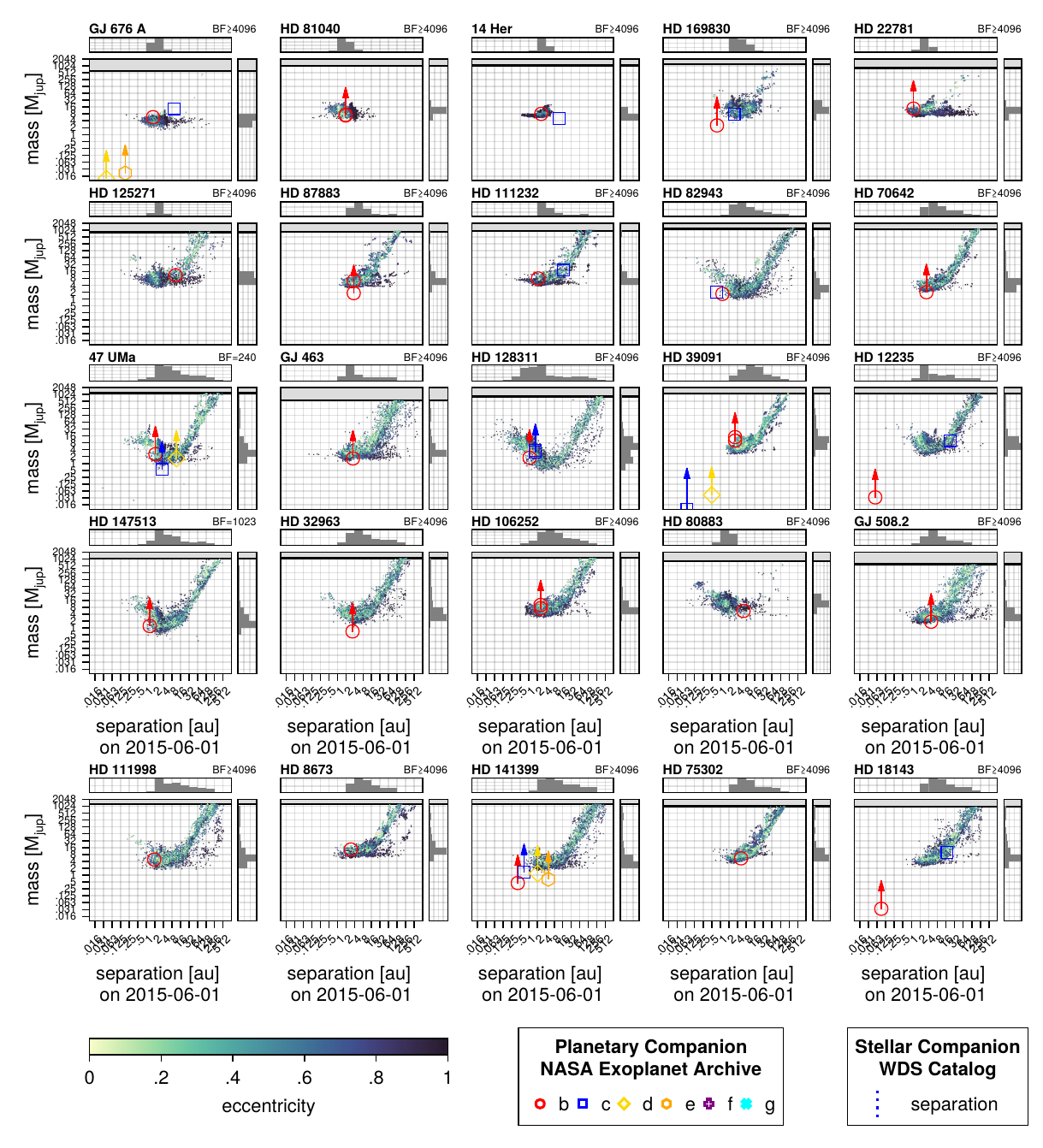}
\caption{Marginal mass-separation posteriors from absolute astrometry for 50 exoplanetary systems hosting Jovian exoplanets. 
The separation calculated from our posteriors is the 3D separation of companion from the star in AU in mid-2015.
The BF label at the top-right of each plot indicates the Bayes factor in favor of a companion. The colored markers indicate known planetary mass companions from the NASA Exoplanet Archive. The color of the marker indicates the planet name. The semi-major axis and mass of each known planet are plotted. In cases where no true mass is available, an arrow indicates the value is $m \sin{i}$ from radial velocity measurements, and the true mass may be higher depending on inclination.}
    \label{fig:nea-stars}
\end{figure*}
\begin{figure*}
    \centering
    \includegraphics[width=\linewidth]{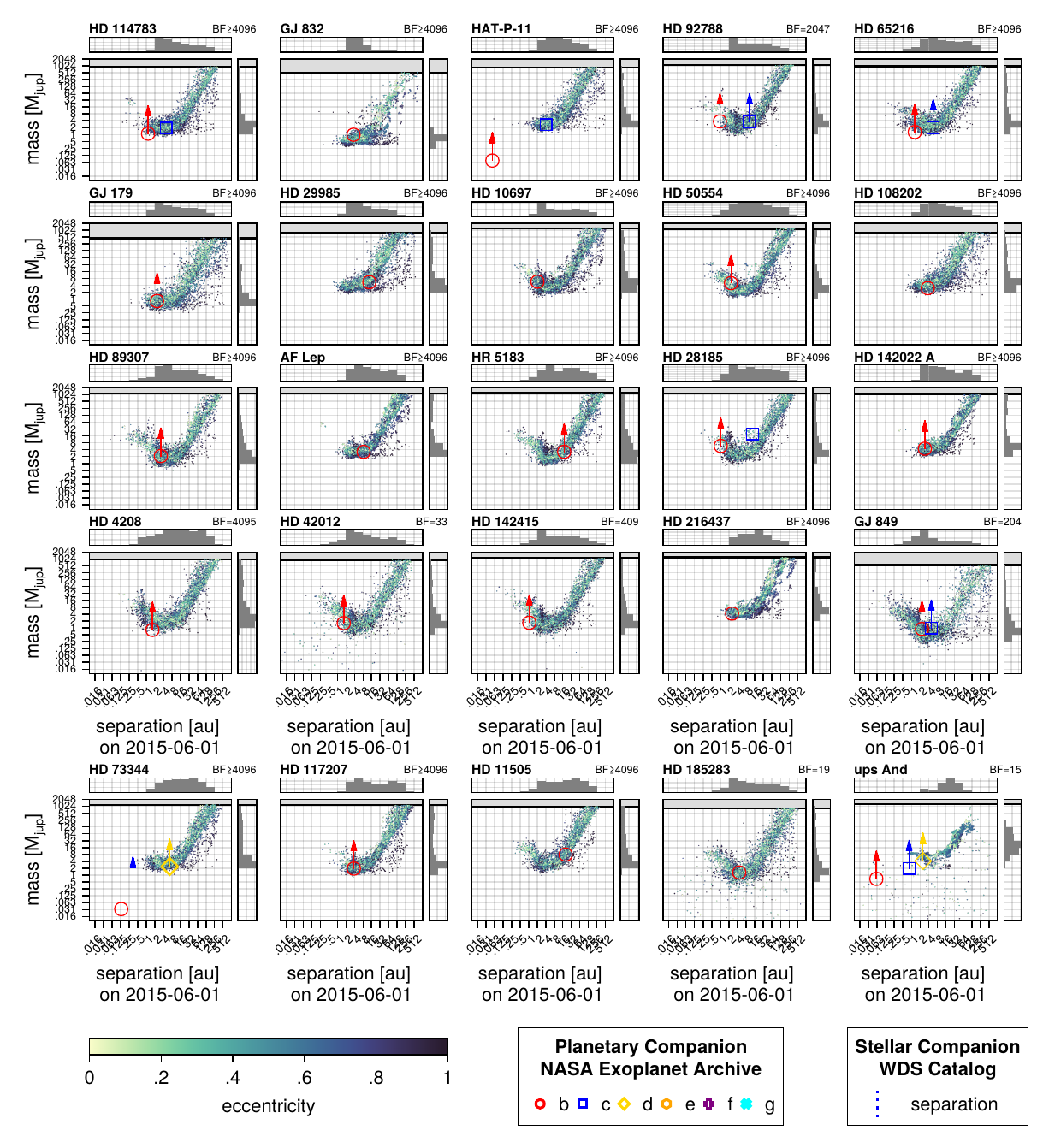}
    \caption{Continued from Figure \ref{fig:nea-stars}.}
    \label{fig:nea-stars-2}
\end{figure*}

\begin{figure*}
    \centering
    \includegraphics[width=\linewidth]{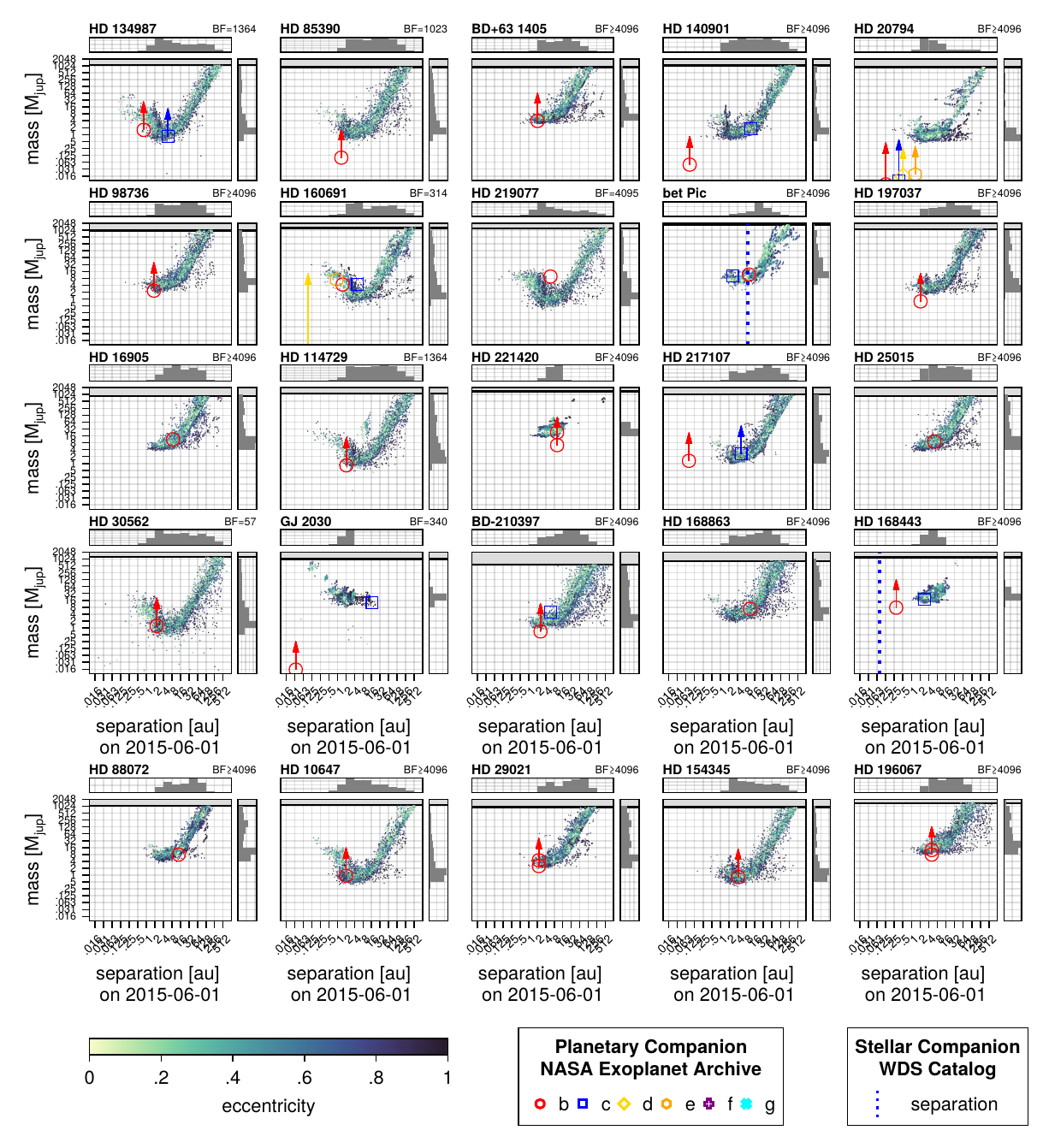}
    \caption{Continued from Figure \ref{fig:nea-stars-2}.}
    \label{fig:nea-stars-3}
\end{figure*}

\begin{figure*}
    \centering
    \includegraphics[width=\linewidth]{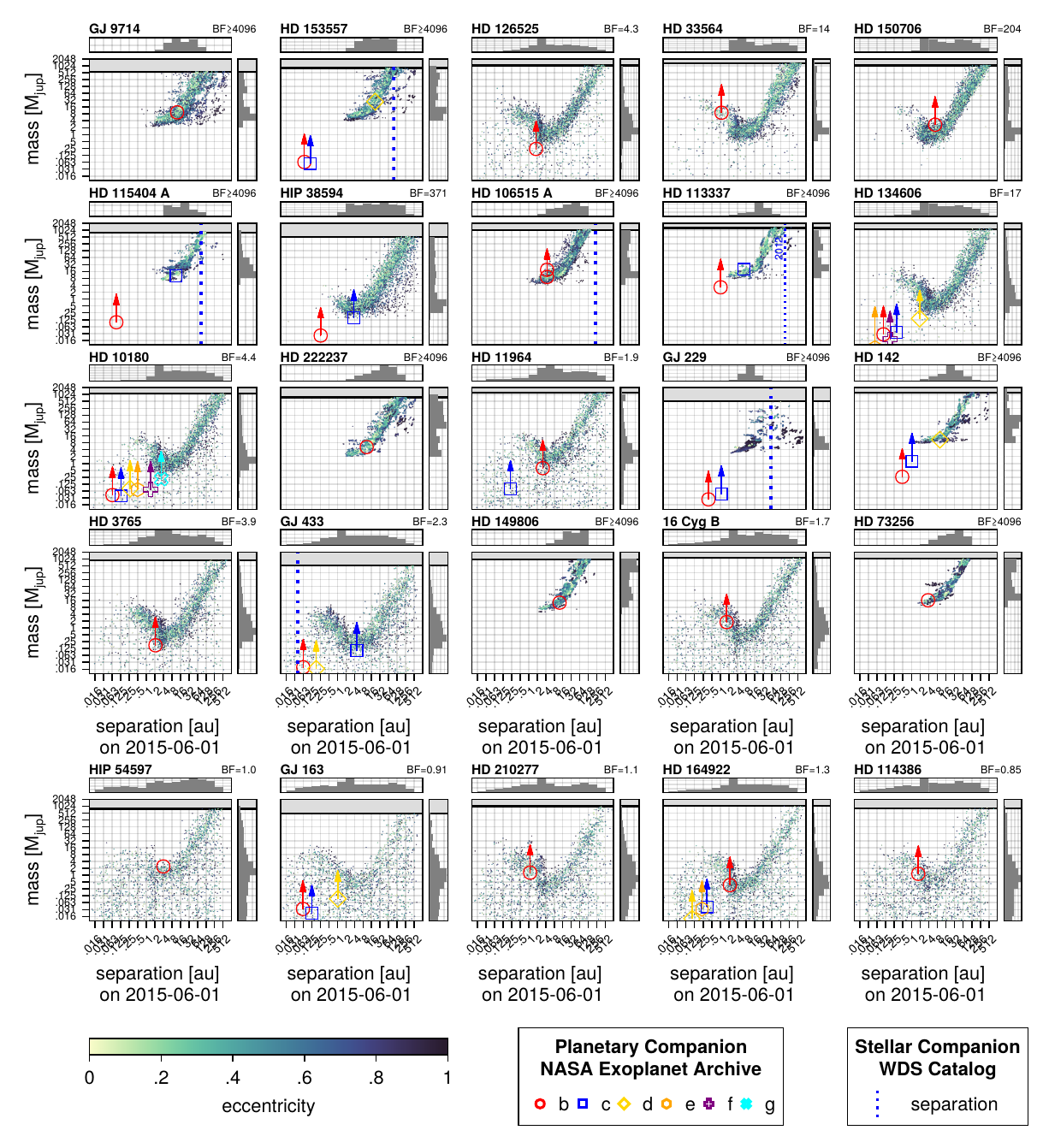}
    \caption{Continued from Figure \ref{fig:nea-stars-3}.}
    \label{fig:nea-stars-4}
\end{figure*}

\begin{figure*}
    \centering
    \includegraphics[width=\linewidth]{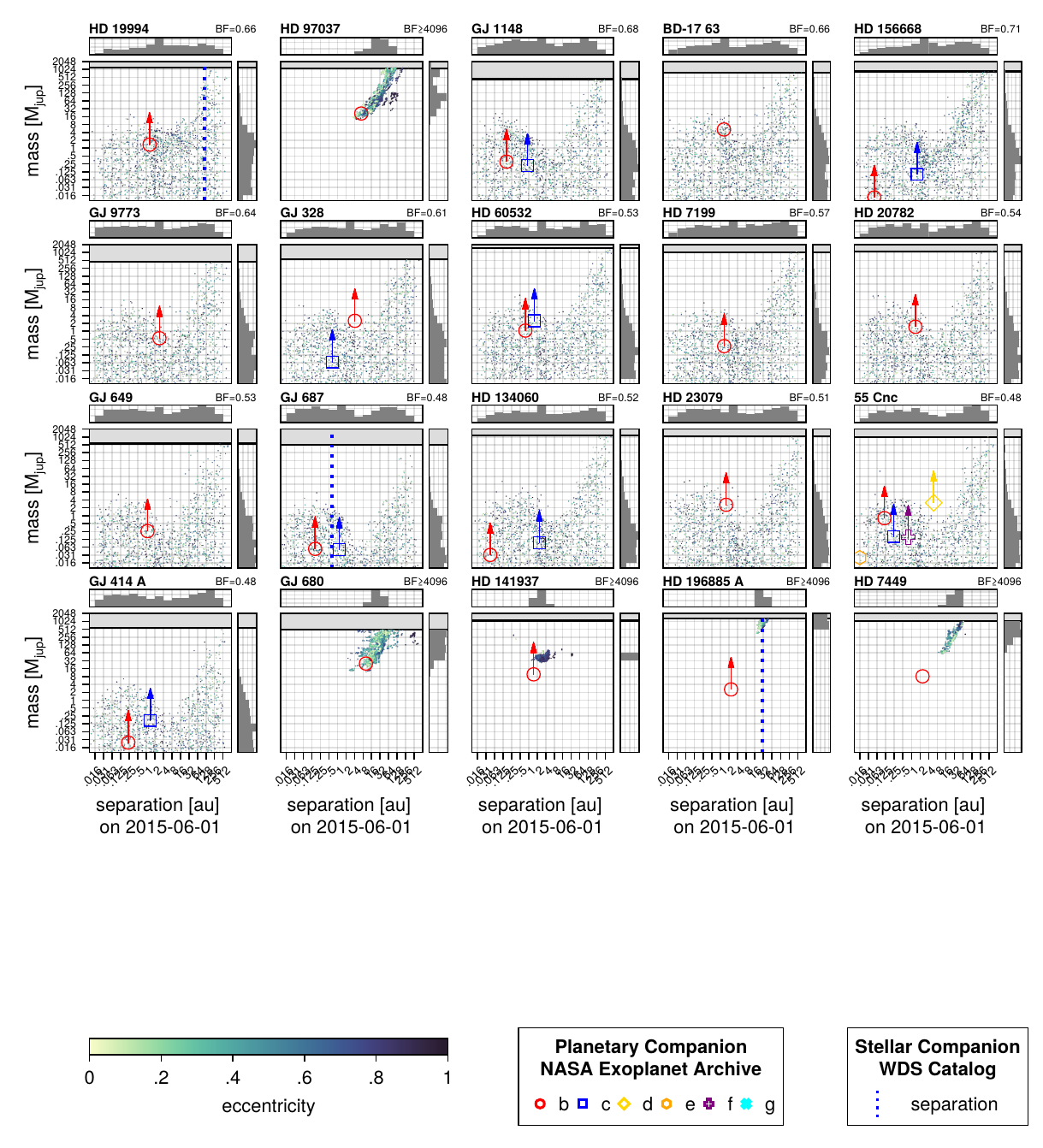}
    \caption{Continued from Figure \ref{fig:nea-stars-4}. The apparent disagreement for HD 141937 is in fact caused by significant orbital inclination not reflected in the NASA Exoplanet Archive solution. The true mass of $27.42^{+6.78}_{-9.86} \; \mathrm{M_{jup}}$ agrees well with our posterior. In the case of HD 196885 A, the astrometric signal from the stellar companion dominates (separation in 2023 indicated by the blue dashed line).}
    \label{fig:nea-stars-5}
\end{figure*}

\subsection{RV Quiet Stars}

In order to assess the rate of false positives, we had the goal of identifying a set of standard stars where any planetary or stellar companion signal could be rejected \emph{a-priori} without looking at Gaia or Hipparcos data. 

We began by finding a set of stars with archival RVs fetched from DACE \citep{dace} spanning 1-2 decades, all showing minimal long term trends, scatter, and having no significant signals with periods greater than 500 days.  We removed stars that had common-proper motion companions nearby in GDR3. 

It turns out, however, that many of these stars still possessed significant HGCA accelerations. This is because the long term Gaia-Hipparcos proper motion anomaly is sensitive to wide companions that are simply not detectable through any other current means, besides perhaps very deep imaging. These are stellar companions that are far enough away so as not to produce a significant RV trend, even over decades, but not bright enough to be detected independently by our Gaia common proper motion checks. 

We instead compromise by finding a set of stars where the available RV data combined with long-term HGCA data rule out any signal probed by the shorter term metrics from the Hipparcos IAD, GDR2 vs.\ GDR3 acceleration, astrometric excess noise, and Gaia RV variability. This does not allow us to estimate an absolute false-positive rate, but does allow us to measure a conditional false-positive rate: if the star does not have a significant HGCA acceleration (the regime we cannot reliably probe through external checks), then the false positive rate for the remaining data (in regimes where we can rule out companions \emph{a-priori} via RV surveys) is less than some value. 

\begin{figure*}
    \centering
    \includegraphics[width=\linewidth]{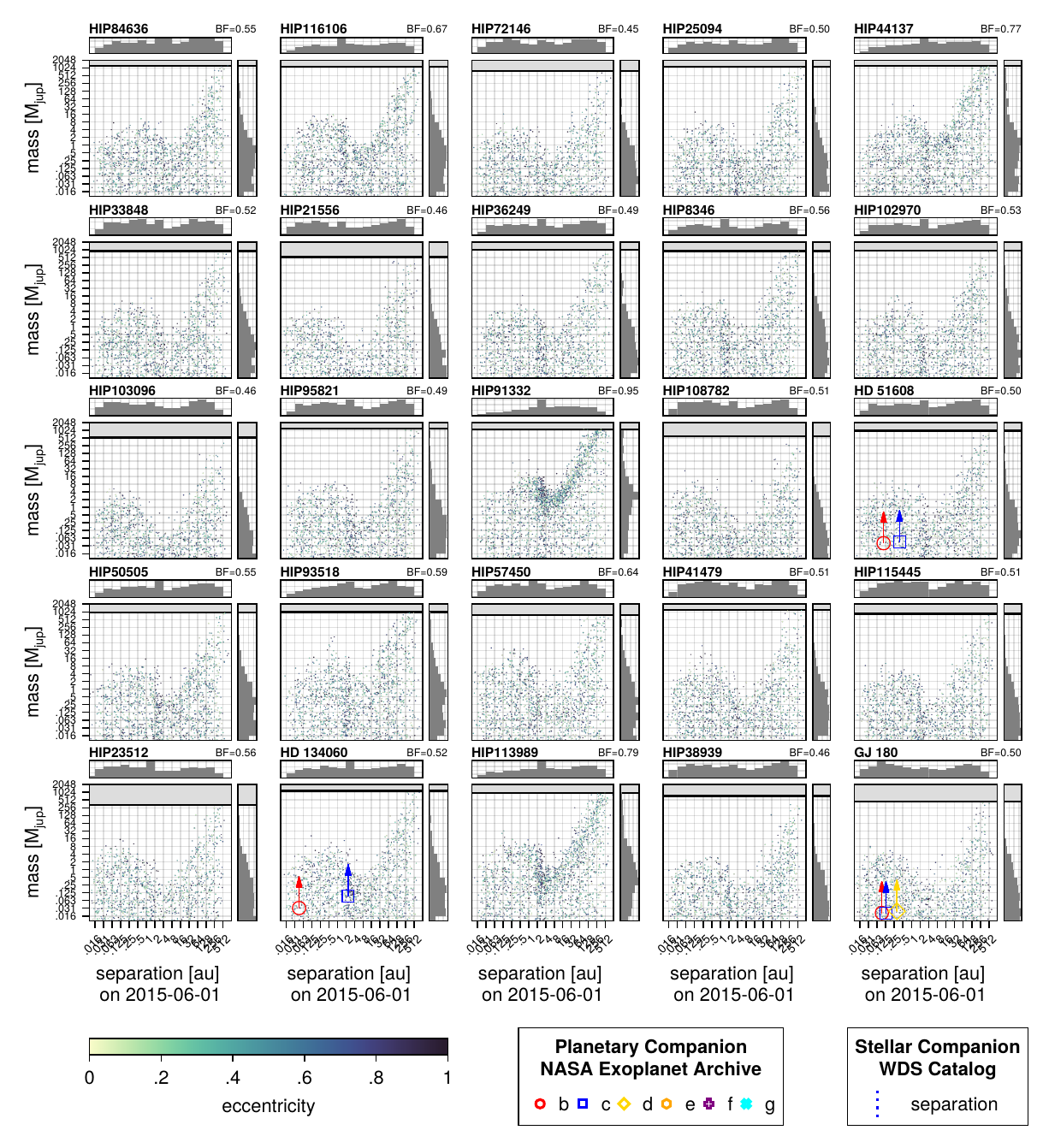}
    \caption{Orbit posteriors for stars with no long-term RV trends or variability on periods greater than 500 days, and no long-term Hipparcos-Gaia proper motion anomaly. These results show that the data used in this work (besides Hipparcos-Gaia) have a low false-positive rate. This does not consider false-alarms from equal flux stellar binaries, which are excluded from this sample by the lack of RV trends. \label{fig:standard-stars}}
    \label{fig:rvstandards}
\end{figure*}

Performing this test, we search a set of 25 RV quiet Hipparcos stars with HGCA proper motion anomaly $\chi^2<1$. These posteriors are presented in Figure \ref{fig:standard-stars}. We find as desired that in all cases the Bayes factor in favor of a companion is less than one, and so can estimate that the rate of false positives---from data other than the HGCA---is better than 1/25.
This does not consider potential astrophysical false-alarms from near-equal flux binaries masquerading as a planets due to flux contamination since no such stars were included in the sample.

\subsection{14 Her Joint RV and Astrometry Model}\label{sec:14-her-rv}

\begin{figure*}
    \centering
    \includegraphics[width=\linewidth]{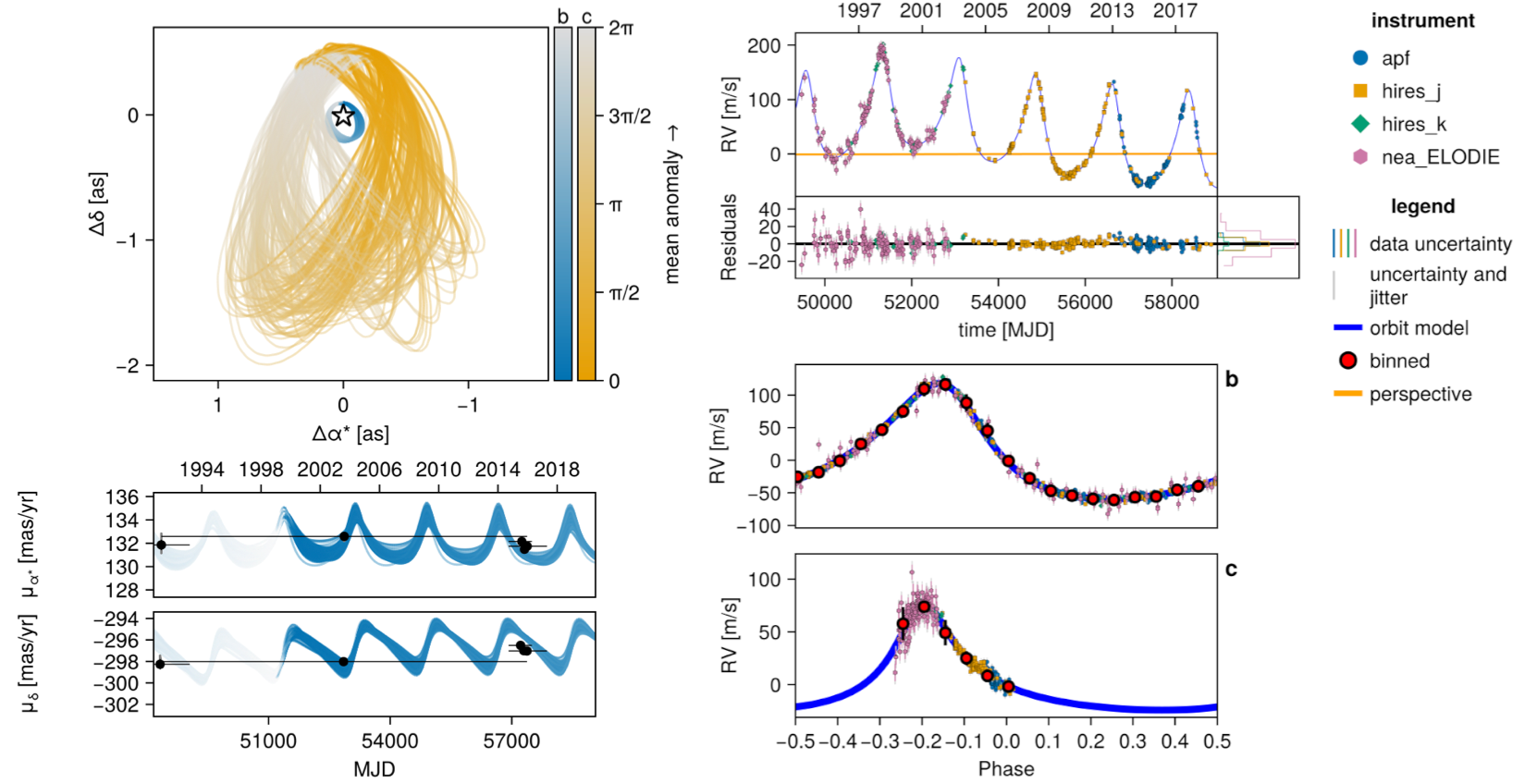}
    \caption{Two-companion model of 14 Her RVs and absolute astrometry illustrating how the additional astrometric data-points can be used to improve constraints on the inclinations and masses of the planets.}
    \label{fig:14-her-2pl-rv}
\end{figure*}

\begin{figure*}
    \centering
    \includegraphics[width=\linewidth]{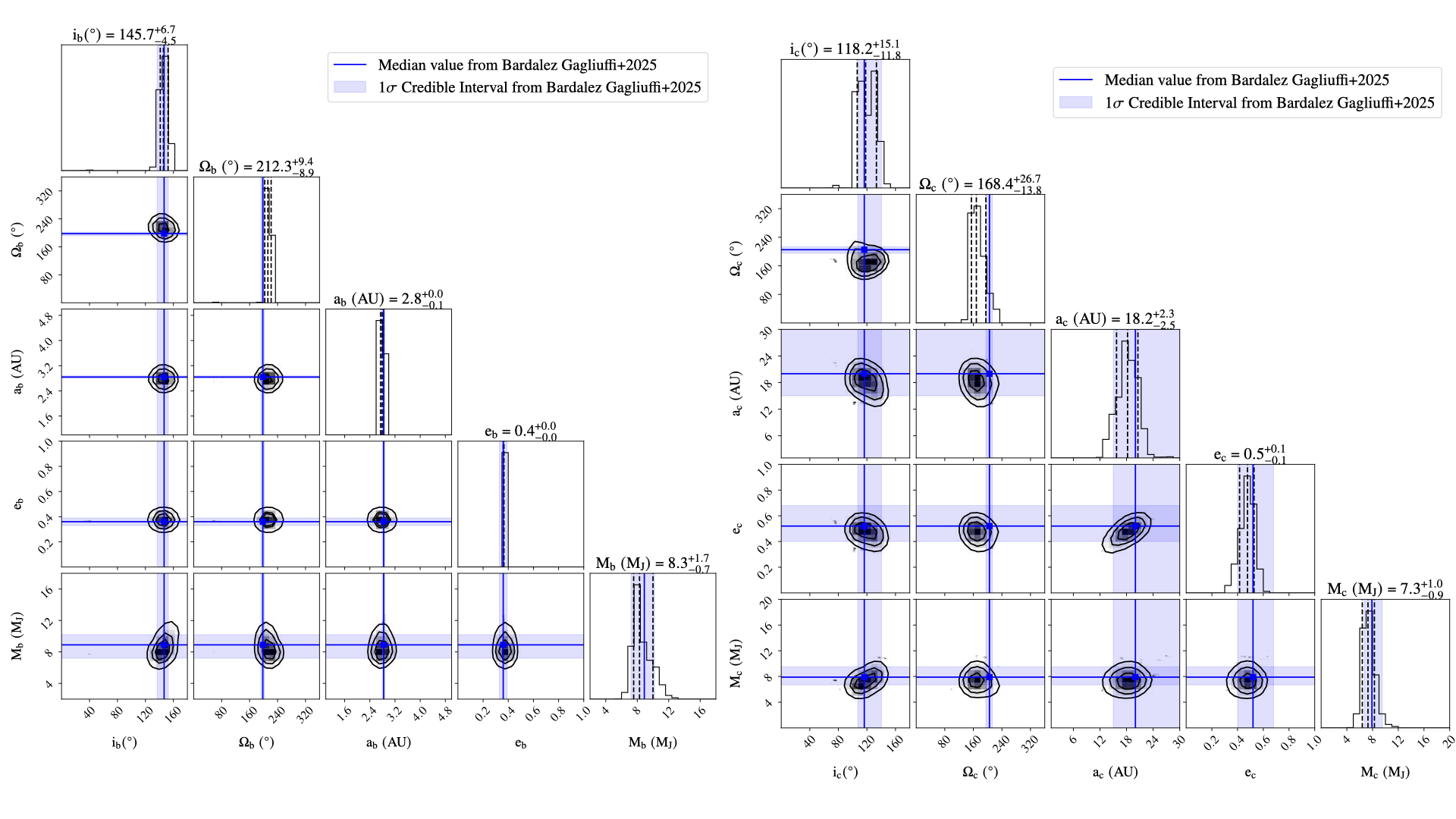}
    \caption{Joint posterior distributions for orbital parameters and true mass of 14 Her b (left) and c (right) from our 2-planet orbital fitting using RVs and absolute astrometry. The blue line and shaded region show the median and $1\sigma$ credible interval for the same parameters from \cite{14herc} which also included the observed position of the planet from JWST image. The ranges are compatible despite us not including the relative astrometry.}
    \label{fig:14-her-2pl-corner}
\end{figure*}

These constraints from Gaia and Hipparcos can of course also be freely combined with external data to further constrain orbits. To demonstrate, we modeled both Gaia and Hipparcos data combined with precision radial velocities.
We neglect companion flux contributions, since both planets are expected to be negligible in the visible band for \emph{Gaia} and \emph{Hipparcos}. We used 418 archival radial velocities of 14 Her from ELODIE \citep{elodie} taken between April 1994 and August 2003, 210 RVs from  APF \citep{apf, Rosenthal2021} taken between January 2014 and April 2019, and 284 RVs from HIRES \citep{hires, Rosenthal2021} taken between April 1997 and February 2020. We consider two planets in our fitting, with seven parameters to describe the orbits and mass of each planet: the orbital period $P$, the mass ratio between planet and host star $q$, the eccentricity $e$, and the argument of periastron $\omega$, the cosine  of its orbital inclination $\cos{i}$, the longitude of the ascending node $\Omega$, the position angle at reference epoch $\theta$. Additional parameters—including noise models for both the RVs and astrometry, RV zero points, and the astrometric epoch selection—are listed in Table~\ref{tab:priors}.

We implemented the two companion model in Octofitter and sampled the posterior using the stabilized variational nonreversible parallel tempering algorithm implemented in \textit{Pigeons.jl} \citep{pigeons}, as before. We sample the model for $2^{14}$ scans. Figure~\ref{fig:14-her-2pl-rv} shows the astrometric measurements and model, along with the RV data and corresponding model. Figure~\ref{fig:14-her-2pl-corner} shows the joint posteriors for the orbital inclination, longitude of the ascending node, semi-major axis, eccentricity, and planet mass of 14 Her b and c. We obtain an orbital inclination of $145^{\circ}.7^{+6.7}_{-4.5}$ for 14 Her b and $118^{\circ}.2^{+15.1}_{-11.8}$ for 14 Her c. Our results are consistent with the values reported by \cite{14herc}, even though we did not include the JWST relative astrometry. \cite{BardalezGagliuffi2021} performed a similar two-planet fit for 14 Her using RVs and the Hipparcos–\emph{Gaia} astrometric acceleration, and constrained the inclination of 14 Her c to $101^{+31}_{-33}$~deg. By incorporating not only the Hipparcos–\emph{Gaia} acceleration, but also the DR2–DR3 proper-motion difference and the UVEA measurements, we significantly tighten the constraint on the orbital inclination of 14 Her c.

\section{Discussion}\label{sec:discussion}

\subsection{Contrast with GaiaPMEX}

This work greatly benefits from using the UEVA approach to calibrating the Gaia RUWE or astrometric excess noise published in \citet{pmex1}. Our work differs first in that we perform an independent calibration of the Gaia astrometric noise by extending their procedure to also estimate uncertainty in noise values as a function of magnitude, color, and position in the sky, and then including this uncertainty in our models.

Our work also differs by the inclusion of our calibrated GDR2 proper motions, GDR3 vs.\ GDR2 scaled position differences, the Hipparcos IAD, and the Gaia RV variability constraints from \texttt{paired} \citep{paired}. As a result, our joint constraints can rule out inner and outer stellar companions in more cases.

In addition, we use the fact that we have modeled the astrometric excess noise imparted by the companion, assuming homoscedasticity of Gaia uncertainties, to scale down the Gaia DR3 estimated parameter  uncertainties to match those expected of a single star. This improves the constraints from DR3 vs.\ Hiparcos and DR3 vs.\ DR2 proper motion anomalies for any stars with a significant UEVA signal.

A direct comparison is possible for HD 81040 highlighted in Figure 20 of \citet{pmex1} and shown near the top-left panel of our Figure \ref{fig:nea-stars}. While the \texttt{GaiaPMEX} approach detects the companion and is able to rule out a wide-separation stellar binary (unlike HGCA data alone), it cannot rule out that the signal could be caused by an inner spectroscopic binary companion. Our full joint posterior rules out an inner stellar companion conclusively and places the companion near the deuterium-burning limit. 

By contrast, for AF Lep (HIP 25486)---one of the other illustrative cases presented in \citet{pmex1}---our improvement is more modest. While HD 81040 b has an orbital period on the order of three years \citep{hd81040}, AF Lep b has an orbital period on the order of 20 years \citep{AFLep2023}. So while the addition of Gaia DR3 vs.\ DR2 proper motion anomaly is able to resolve curvature from the orbit of HD 81040 b, the additional information is less constraining for the longer period orbit of AF Lep. Our additional constraints assign a $60\%$ posterior probability that the companion is $13.7 \, \mjup$ or below, but cannot conclusively confirm the companion is planetary-massed without external data.

\begin{figure}
    \centering
    \begin{adjustwidth}{-1.cm}{-0.8cm} 
    \includegraphics[width=\linewidth]{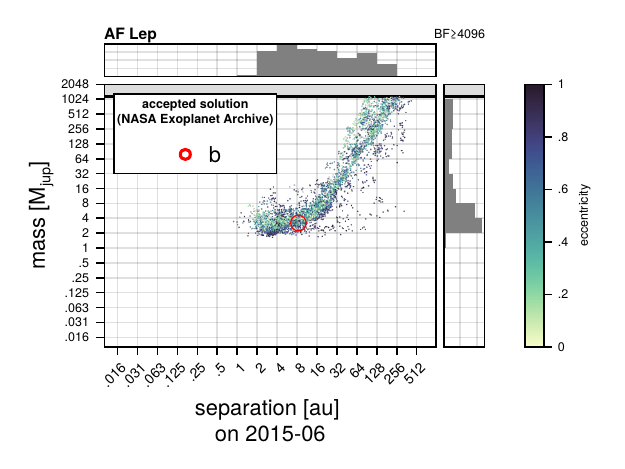}
    \end{adjustwidth}
    \caption{Mass vs.\ separation marginal posterior for AF Lep (HIP 25486). In this case, the orbital period of planet b is long enough not to be well-constrained by the GDR3-GDR2 proper motion differences, and so the posterior is similar to that presented in \citet{pmex1}. The small points represent draws from the posterior, colored by eccentricity. The horizontal axis is the 3D separation of companion from the star in AU in mid-2015. The larger colored markers present the mass and semi-major axes of the planet as retrieved from the accepted entry in the NASA Exoplanet Archive. }
    \label{fig:aflep}
\end{figure}

A final notable difference from the \citet{pmex1} paper series is that in our models we explicitly do not consider the flux of the secondary companion. Those works use a semi-empirical mass-luminosity relation from \citet{kiefer_2020} assuming an age of 5 Gyr. This is an important consideration because equal-flux binaries can wash out the amplitude of photocentre motion detected by Gaia and Hipparcos, creating an astrometric signal that matches that of an exoplanet. As previously discussed, we leave detection of false alarms from these equal flux binaries to a future work, and for now model only the ``dark mass.'' That said, modeling the flux of the secondary is supported by Octofitter and any mass-luminosity model can be inputted by users.

\subsection{\textit{Caveat Emptor}}\label{caveats}

We now warn potential users of several important caveats to the use of this catalog.

These known shortcomings include the following.
First, the correlation derived between the GDR3 and GDR2 proper motions (Eq. \ref{eq:rho}) is only a rough approximation. In actuality, the correlation between these measurements is complex and dependent on the scan angles and measurement epochs. 

Second, it should be noted that the correlations between the R.A. and Dec.\ values of the DR3--DR2 scaled position difference are not considered.
Similarly, the correlation between the DR3--DR2 scaled position difference and either the DR2 or DR3 proper motions is not considered.
Adding these correlations would serve to increase the significance of observed discrepancy, but would require recalculating the error inflation factors in the catalog.

In addition, constraints from the Gaia astrometric excess noise may be correlated with the change in proper motion between Gaia DR3 and DR2. It is not clear at this time how such correlation should be modeled. Consider for instance a short period planet with orbital period much less than the time baseline between the central epochs of DR2 and DR3 could result in significant astrometric excess noise, but have no impact on the net change in proper motion between DR2 and DR3. In other cases where the period is comparable to this baseline, the astrometric excess noise must be correlated with the amplitude (though not the direction) of the acceleration.
It may be that this unmodeled correlation results in overconfidence in some cases.
Without an analytic expression for these correlations, we instead argue that the empirical results indicate these limitations are not significant in practice. Future work could refine these details and improve the constraints provided, and Gaia DR4 epoch astrometry will render the issue moot.

Another issue is that this catalog is likely to be inaccurate for bright stars--- especially those brighter than 4th magnitude---because Gaia's bright star processing changed significantly between DR2 and DR3 and due to the small sample size of these bright stars which limits our ability to calibrate these systematics.

As a final word of caution, users should be aware that the radial velocity column included in this table is taken entirely from Gaia DR3. Other sources of radial velocity information (including those used within the HGCA) may be more reliable for some targets. Having access to the most reliable radial velocity of the system is important for modeling secular acceleration, which can masquerade as a long-period drift or perturbation. For systems where the radial velocity is not reliable, we encourage users to marginalize over the uncertainty in the system's RV by adding it as an additional parameter to their model.

\section{Conclusion}\label{sec:conclusion}

We have presented a catalog that calibrates Gaia DR2 positions and proper motions against Gaia DR3 and combines this data with the HGCA \citep{hgca_dr3}, Gaia astrometric uncertainty calibration values \citep[calculated independently following a similar procedure to ][]{pmex1}, and Gaia RV uncertainty calibration values from \citet{paired}.

This catalog can be used to add additional orbit constraints to joint imaging/RV/astrometry fits or an independent indicator a star may host a companion.

We develop an extension to the Bayesian orbit modeling code Octofitter that jointly models:
\begin{itemize}
    \item The Hipparcos IAD,
    \item Hipparcos positions and proper motions,
    \item Gaia DR2 positions and proper motions,
    \item Gaia DR3 positions and proper motions,
    \item Gaia astrometric excess noise,
    \item Gaia radial velocity excess noise,
\end{itemize}
to provide the tightest constraints yet on the existence of stellar and planetary companions using only Gaia and Hipparcos.

This code is made public in a new release of Octofitter (v8), and the combined catalog is made available at \href{http://dx.doi.org/10.11570/26.0002}{http://dx.doi.org/10.11570/26.0002}.

We demonstrated the approach on a set of 25 binary stars with well determined orbits, known exoplanetary systems from the NASA Exoplanet Archive, and a set of stars with no long-term RV or Gaia-Hipparcos trends, finding good agreement with published solutions and no detected false-positives that could be independently ruled out with archival RV data.

\begin{acknowledgments}

This work has made use of data from the European Space Agency (ESA) mission Gaia (\href{https://www.cosmos.esa.int/gaia}{https://www.cosmos.esa.int/gaia}), processed by the Gaia Data Processing and Analysis Consortium (DPAC), \href{https://www.cosmos.esa.int/web/gaia/dpac/consortium}{https://www.cosmos.esa.int/web/gaia/dpac/consortium.}
This work made use of the Gaia Observation Forecast Tool (GOST).

This research used the facilities of the Canadian Astronomy Data Centre operated by the National Research Council of Canada with the support of the Canadian Space Agency.

This research was enabled in part by support provided by the Digital Research Alliance of Canada (alliancecan.ca).

This research has made use of the NASA Exoplanet Archive, which is operated by the California Institute of Technology, under contract with the National Aeronautics and Space Administration under the Exoplanet Exploration Program.

J.S. acknowledges financial support from the Natural Sciences and Engineering Research Council of Canada (NSERC) through the Canada Graduate Scholarships Doctoral (CGS D) program, and from the Heising–Simons Foundation through the 51 Pegasi b Fellowship (grant \#2025-5885).
J.W.X is thankful for support from the Heising-Simons Foundation 51 Pegasi b Fellowship (grant \#2025-5887).

This work benefited from the 2025 Exoplanet Summer Program in the Other Worlds Laboratory (OWL) at the University of California, Santa Cruz, a program funded by the Heising-Simons Foundation and NASA.
\end{acknowledgments}

\facility{Gaia, Hipparcos, Keck:HIRES, Lick:APF, OHP:ELODIE, TESS, Exoplanet Archive}

\bibliography{ms}{}
\bibliographystyle{aasjournalv7}

\end{document}